\PassOptionsToPackage{pdfpagelabels=false, pageanchor=false, hyperfootnotes=false, plainpages=false,hypertexnames=false}{hyperref} 
\documentclass[fleqn,usenatbib]{mnras}
\usepackage[T1]{fontenc}
\usepackage{ae,aecompl}
\usepackage{graphicx}
\usepackage{amsmath, bm}
\usepackage{amssymb}
\usepackage{txfonts}
\usepackage{mathptmx}

\usepackage{textcomp}
\usepackage{gensymb}
\usepackage{etoolbox}
\usepackage{xspace}
\usepackage{paralist}
\usepackage{blindtext}
\usepackage[english]{babel}
\usepackage[usenames,dvipsnames, table]{xcolor}

\hypersetup{colorlinks,linkcolor={NavyBlue},citecolor={NavyBlue},urlcolor={NavyBlue}}

\newtoggle{changed}
\toggletrue{changed}

\renewcommand{\vec}[1]{\boldsymbol{#1}}
\newcommand{\pc}{\,\rm{pc}\xspace}
\newcommand{\kpc}{\,\rm{kpc}\xspace}
\newcommand{\kms}{\,\rm{km\,s^{-1}}\xspace}
\newcommand{\kmskpc}{\,\rm{km\,s^{-1}\,kpc^{-1}}\xspace}
\mathchardef\mhyphen="2D

\newcommand{\nmagic}{NMAGIC\xspace}
\newcommand{\brava}{BRAVA\xspace}
\newcommand{\argos}{ARGOS\xspace}
\newcommand{\twomass}{2MASS\xspace}
\newcommand{\ukidss}{UKIDSS\xspace}

\newcommand{\ogle}{OGLE\xspace}
\newcommand{\vvv}{VVV\xspace}
\newcommand{\apogee}{APOGEE\xspace}
\newcommand{\gibs}{GIBS\xspace}

\newcommand{\Msun}{\,\rm{M}_{\odot}\xspace}

\newcommand{\Gyr}{\,\rm{Gyr}}

\newcommand{\Feeq}{\textit{M}}
\newcommand{\Fe}{\textrm{[Fe/H]}}
\newcommand{\dex}{\textrm{dex}\xspace}

\title[Chemodynamical Modelling of the Galactic Bulge and Bar]{Chemodynamical Modelling of the Galactic Bulge and Bar}

\author[M. Portail et al.]
  {Matthieu~Portail$^1$\thanks{Email: \{portail, wegg, gerhard\}@mpe.mpg.de}, Christopher~Wegg$^1$, Ortwin~Gerhard$^1$ and Melissa~Ness$^2$\\
  $^1$ Max-Planck-Institut f\"{u}r Extraterrestrische Physik, Gie\ss enbachstra\ss e, D-85741 Garching, Germany\\
  $^2$ Max-Planck-Institut  f\"{u}r  Astronomie,  K\"{o}nigstuhl  17, D-69117 Heidelberg, Germany}

\date{Accepted 2017 May 23. Received 2017 May 22; in original form 2016 December 15}

\pubyear{2017}
\begin{document}
\phantomsection\label{firstpage}
\pagerange{\pageref{firstpage}--\pageref{lastpage}}
\maketitle

\begin{abstract}
We present the first self-consistent chemodynamical model fitted to reproduce data for the galactic bulge, bar and inner disk. We extend the Made-to-Measure method to an augmented phase-space including the metallicity of stars, and show its first application to the bar region of the Milky Way. Using data from the \argos and \apogee (DR12) surveys, we adapt the recent dynamical model from Portail et al. to reproduce the observed spatial and kinematic variations as a function of metallicity, thus allowing the detailed study of the 3D density distributions, kinematics and orbital structure of stars in different metallicity bins. We find that metal-rich stars with $\Fe \geq -0.5$ are strongly barred and have dynamical properties that are consistent with a common disk origin. Metal-poor stars with $\Fe \leq -0.5$ show strong kinematic variations with metallicity, indicating varying contributions from the underlying stellar populations. Outside the central $\kpc$, metal-poor stars are found to have the density and kinematics of a thick disk while in the inner $\kpc$, evidence for an extra concentration of metal-poor stars is found. Finally, the combined orbit distributions of all metallicities in the model naturally reproduce the observed vertex deviations in the bulge. This paper demonstrates the power of Made-to-Measure chemodynamical models, that when extended to other chemical dimensions will be very powerful tools to maximize the information obtained from large spectroscopic surveys such as APOGEE, GALAH and MOONS.
\end{abstract}


\section{Introduction}
\label{section:introduction}

In the last two decades, our understanding of the inner Galaxy and in particular its central bulge has dramatically changed, driven by the availability of large-scale surveys. The galactic bulge was initially thought to be a classical bulge, spheroidal remnant of mergers that happened during the hierarchical assembly of the Galaxy. This was supported by its old age \citep{Ortolani1995} and the presence of a vertical metallicity gradient, which was thought at the time to rule out other formation scenarios \citep{Friedli1994}. The picture changed when non-axisymmetries, found consistently in the H{\scriptsize I} and CO gas flow \citep{Binney1991, Englmaier1999, Fux1999}, in the near-infrared light distribution \citep{Blitz1991,Weiland1994, Bissantz2002} and star counts \citep{Nakada1991,Stanek1997, Lopez-Corredoira2005} revealed the presence of a triaxial and rather short boxy bar structure in the bulge \citep{Dwek1995, Binney1997}, extending to $2-3\kpc$ from the centre. 

Recently, \citet{Nataf2010} and \citet{McWilliam2010} independently discovered the so-called split red clump, caused by the bimodal density of red clump giants (RCGs) in the bulge. This bimodality is best illustrated by the work of \citet{Wegg2013} who were able to reconstruct the 3D density of 8 million RCGs stars identified in the \vvv catalogue, by direct deconvolution of their magnitude distributions in about 300 fields throughout the bulge. Their density map reveals very clearly the so-called boxy/peanut (B/P) shape of the galactic bulge. Such B/P structured bulges had already been seen in N-body simulations of barred stellar disks and indicated a different formation scenario, where the galactic bulge formed from disk stars through the vertical instability of a stellar bar \citep{Combes1981, Raha1991, Debattista2000, Athanassoula2005, Martinez-Valpuesta2006}. 

Additional pieces of evidence now support this formation scenario for the galactic bulge. First, spectroscopic surveys such as the \brava \citep{Kunder2012} and \argos \citep{Freeman2013} surveys showed that the bulge rotates cylindrically, characteristic of a disk origin. Then, \citet{Wegg2015} showed by combining the \vvv, \ukidss and \twomass catalogues that the galactic bulge smoothly segues into the flat long bar, that extends to about $5\kpc$ from the Galactic Centre. They find that both the bulge and the long bar appear at a consistent angle with respect to the line of sight towards the Galactic Centre, with no need for a secondary galactic bar component, as proposed earlier (\citealt{Benjamin2005}; \citealt{Lopez-Corredoira2005}; \citealt{Cabrera-Lavers2008}; but see also \citealt{Martinez-Valpuesta2011}). Henceforth, the galactic bulge is now believed to be the vertically extended part of a buckled long bar component, qualitatively similar to the bar produced in N-body simulations. 

The first comprehensive and self-consistent dynamical model of the bulge and bar structure in the Milky Way was constructed recently by \citet[][hereafter \hyperlink{P17}{P17}]{Portail2016a}. Using the Made-to-Measure (M2M) method, \hyperlink{P17}{P17} were able to find a dynamical equilibrium model that reproduces simultaneously the RCG density in the bulge and bar region from a combination of the \vvv, \ukidss and \twomass surveys together with stellar kinematics from the \brava, \ogle and \argos surveys.

Based on our new knowledge of the current dynamical state of the inner Milky Way, the next stage is to try to understand its chemodynamical formation history. Although individual orbits of stars are scrambled during the bar formation and subsequent buckling instability, \citet{Martinez-Valpuesta2013} showed that the Jacobi energy of stars is largely conserved during the process. Thus, by advanced modelling of the current state of the Galaxy we can hope to trace back earlier stages \citep{DiMatteo2014}. This is challenging and requires as much data as possible, in particular chemical information on stars. Several studies have tried to infer a formation scenario for the bulge based on measurements of the stellar metallicity and its gradient along the bulge minor axis. \citet{Martinez-Valpuesta2013} showed that the presence of a vertical metallicity gradient in the bulge was in itself not a strong argument for the presence of a classical bulge component as previously thought. They found that an initial radial gradient in a disk can, after bar formation and buckling, result in a vertical metallicity gradient similar to that observed in the galactic bulge \citep{Minniti1995, Zoccali2008, Gonzalez2011}, thus suggesting that the galactic bulge could have a pure disk origin. However, other studies \citep[e.g.][]{Babusiaux2010, Gonzalez2011a, Hill2011, Uttenthaler2012,Rojas-Arriagada2014,Zoccali2016} have found evidence for two stellar populations in the bulge with different kinematics: a metal-rich population centred on $\Fe\sim0.25$, rapidly rotating and dynamically cold, and a metal-poor population centred on $\Fe\sim-0.3$, dynamically hotter and more slowly rotating. These authors associated the metal-rich component to a B/P bulge with disk origin and suggested that the metal-poor component is an old classical spheroid, which in Baade's window would account for about half of the stars in the bulge.

This view of a prominent classical bulge in the Galaxy has been challenged by the \argos survey and its associated series of papers. Using the full \argos sample, \citet{Ness2013a} showed that the metallicity distributions of stars in different fields across the inner Galaxy were well represented by five different components whose relative fractions change from field to field, thus explaining the vertical metallicity gradient of the bulge as a change in the relative fractions of the components. For each of these components, labeled from A to E in the order of decreasing metallicity they speculate a different origin as summarized below:
\begin{itemize}
  \item A $(\Fe\sim0.1)$: Metal-rich cold component of the B/P bulge, concentrated to the plane and associated to a more recent star formation episode in the early disk than B.
  \item B $(\Fe\sim-0.3)$: Main component of the B/P bulge originating from the thin disk before the bar formation and secular evolution.
  \item C $(\Fe\sim-0.7)$: More metal-poor thicker disk-like population that does not significantly support the B/P shape of the bulge. C is associated to the hot and less dynamically responsive inner thick disk at the time when the bar formed.
  \item D $(\Fe\sim-1.2)$: Metal weak thick disk, similar to the thick disk seen in the solar neighbourhood.
  \item E $(\Fe\sim-1.7)$: Very metal-poor stars, associated to either a small classical bulge component or the inner part of the stellar halo.
\end{itemize}

The interpretation of the origin of the different \argos components has been discussed by \citet{DiMatteo2014} and \citet{DiMatteo2015} who concluded from analysing the formation of B/P bulges in N-body simulations that A and B have both a thin disk origin, with B formed on average at larger radii than A while C is likely to be the old thick disk already in place at the time of bar formation. However, this latter conclusion has been questioned by \citet{Debattista2016} who showed recently that co-spatial disk stellar populations with different radial dispersions can be separated by the bar formation into different orbit distributions.

Stars with $\Fe\leq-1$, such as traced by the RRLyrae stars in the bulge, also have a controversial origin. Bulge RRLyrae have been found by \citet{Pietrukowicz2015} to be bar shaped and centrally concentrated (but see also \citealt{Dekany2013} who find instead that they form a spheroid). From spectroscopic measurements, \citet{Kunder2016} found that they form a hot population, with a velocity dispersion of $\sim130\kms$ and rotate only slowly or not at all. They conclude that bulge RRLyrae stars are consistent with forming a small classical bulge component. Instead \citet{Perez-Villegas2016} showed that both their density and kinematics are consistent with what is expected for the inner part of the stellar halo that co-evolved with the galactic bar.

In this context of controversies, the goal of this paper is two-fold. First, we present an extension of the M2M method for constructing chemodynamical models of galaxies by attaching chemical variables to N-body orbits in a self-consistent dynamical model. Such M2M chemodynamical models have the power to extract maximum information from the data and will be a valuable tool in the future for modelling the data provided by ongoing and future large spectroscopic surveys in the Galaxy. We focus here on the metallicity distribution but future applications could extend the modelling to other chemical quantities, such as for example $[\alpha/\rm{Fe}]$. This is described in \autoref{section:chemodynamicalModelling}. Then we investigate what currently available data can tell us about the present state of the inner Galaxy in the seven-dimensional extended phase-space of position, velocity and metallicity.
We use the spatial and kinematic variations of the metallicity of stars seen in the \argos and \apogee DR12 data to construct an M2M chemodynamical model of the galactic bar region. We base the underlying dynamics on the model of \hyperlink{P17}{P17} which is a very good representation of the inner Galaxy and provides us with a library of N-body orbits on to which we attach and adjust a metallicity distribution. We describe this fitting procedure in \autoref{section:data} and show in Sections \ref{section:spatialVariations} and \ref{section:kinematicVariations} the spatial distribution and kinematics of stars in the inner Galaxy for different metallicities. We finally discuss possible formation scenarios for the bulge, bar and inner disk, based on our chemodynamical model and in the light of other works and conclude in \autoref{section:discussion}.

\section{Made-to-measure chemodynamical modelling}
\label{section:chemodynamicalModelling}

\subsection{The Made-to-Measure method}
\label{section:M2Mintro}

The M2M method  is a particle-based modelling technique that allows the creation of constrained equilibrium models \citep{Syer1996}. Initially developed to tailor N-body initial conditions \citep{Syer1996, Dehnen2009}, the M2M method was adapted by \citet{DeLorenzi2007} to create dynamical models of real galaxies by fitting observational data. The idea behind the M2M method is to slowly adapt the particle weights of a N-body model in order to reproduce a given set of data. The N-body weights, representing mass elements in phase-space in the classical N-body approach are simultaneously seen as weighting factors of the N-body orbits on which the particles are. The M2M method has the advantage of being applicable to complex systems like barred galaxies where other modelling techniques such as moment-based methods \citep{Binney1990, Cappellari2009}, classical distribution function-based methods \citep{Dejonghe1984, Qian1994}, action-based methods \citep{Binney2010, Sanders2013} or orbit-based methods \citep{Schwarzschild1979, Thomas2009} would be either not applicable or not currently practical. 

Formally, we represent the equilibrium state of the Galaxy by its distribution function $f(\vec{x, v})$ and write any observable $y$ as
\begin{equation}
\label{equation:generalObservable}
 y = \int K(\vec{x, v}) f(\vec{x, v}) \,d^3\vec{x}\, d^3\vec{v}
\end{equation}
where $\vec{x, v}$ is the phase-space vector and $K$ is called the kernel of the observable. In the M2M method, $f(\vec{x, v})$ is approximated by a discrete sample of $N$ particles with particle weights $w_i(t)$. \autoref{equation:generalObservable} is then evaluated by
\begin{equation}
\label{equation:observable}
 y(t) = \sum_{i=1}^N w_i(t) \times K(\vec{x}_i(t), \vec{v}_i(t)) 
\end{equation}
where $(\vec{x}_i(t), \vec{v}_i(t))$ are the phase-space coordinates of particle $i$ at time $t$. In practice, $y(t)$ is often noisy due to the limited number of particles in the model. A convenient way to significantly reduce the particle noise is to replace the observable $y(t)$ by the temporally smoothed observable $\tilde{y}(t)$, defined as
\begin{equation}
\label{equation:smoothobservable}
\tilde{y}(t) = \int_{\tau} y(t-\tau)\times e^{-\alpha \tau}\, d\tau
\end{equation}
where $\alpha$ sets the temporal smoothing time-scale.

The M2M method consists of adjusting the particle weights $w_i$ in order to maximize a given profit function $F$. This is done through a simple gradient ascent method where the particle weights evolve with time according to
\begin{equation}
  \label{equation:FOC1}
  \frac{dw_i}{dt} = \varepsilon w_i \frac{\partial F}{\partial w_i}
\end{equation}
with $\varepsilon$ a numerical factor that sets the typical time-scale of the weight evolution. The profit function $F$ usually consists of a $\chi^2$ term that drives the model towards some data plus possibly a regularization term for minimizing the broadening of the weight distribution. After a decade of refinement, this modelling technique is now mature and has been used in both the extragalactic \citep[e.g.][]{DeLorenzi2008, DeLorenzi2009, Das2011, Morganti2013, Zhu2014} and Galactic context (e.g. \citealt{Bissantz2004}; \citealt{Long2013}; \citealt{Hunt2014}; \citealt{Portail2015a}; \hyperlink{P17}{P17}). Pioneering work in M2M chemodynamical modelling has also been done recently by \citet{Long2016} who modelled absorption line strengths in four elliptical galaxies.

\subsection{Modelling the metallicity of stars}
\label{section:particleMDF}

In regular M2M modelling, $f$ is simply a function of phase-space. The method can be extended to chemodynamical models by extending the phase-space to chemical dimensions. Introducing the metallicity $\Feeq$ into the phase-space, \autoref{equation:observable} becomes
\begin{equation}
\label{equation:generalObservableZ}
 y = \int_{\Feeq} \int_{\vec{x, v}} K(\vec{x, v}, \Feeq) \,  f(\vec{x, v}, \Feeq) \, d^3\vec{x} \, d^3\vec{v} \, d\Feeq.
\end{equation}
In all the following, we use the term `metallicity' to mean the iron abundance of a star, denoted alternatively by $\Fe$ or shorter $\Feeq$ for readability of the equations.
Several choices can be made in constructing a particle-based representation of $f(\vec{x, v}, \Feeq)$. One possibility is to assign a single metallicity value $\Feeq_i$ to each particle, and approximate $f$ as the sum of delta functions $f(\vec{x, v}, \Feeq) = \sum_i w_i \times \delta(\vec{x}-\vec{x}_i)\, \delta(\vec{v}-\vec{v}_i)\, \delta(\Feeq-\Feeq_i)$. However, in the overall potential, the particles represent orbits that can in reality be populated by many stars of different metallicities. In such a description, modelling a wide distribution of metallicity populating near-identical orbits requires many particles, with equivalent phase-space coordinates but different metallicities. Instead, it is more efficient to adopt for each particle a metallicity distribution function (MDF), denoted $f_{\Feeq, i}$, and write the distribution function as 
\begin{equation}
\label{equation:DF}
 f(\vec{x, v}, \Feeq) = \sum_i w_i \times \delta(\vec{x}-\vec{x}_i) \, \delta(\vec{v}-\vec{v}_i) \, f_{\Feeq, i}(\Feeq).
\end{equation}
The space of all possible MDFs is of infinite dimension and has to be approximated by a finite dimensional space. We model $f_{\Feeq, i}(\Feeq)$ as an expansion on some mathematical set of elementary metallicity distribution functions (EMDFs), common for all particles, and thus represent $f_{\Feeq, i}(\Feeq)$ by a discrete set of numbers. The choice of the number and form of the EMDFs is constrained by both physical and technical considerations. First, as discussed later in \autoref{section:argos}, stars with different metallicities have different luminosity functions and thus usually have a different probability to be detected in a given survey depending on its selection criteria. The set of EMDFs should allow a faithful representation of the full MDF of stars in the Galaxy, and not for example only in a limited metallicity range. Secondly, the number of parameters used to represent the particle MDFs should be large enough to capture all relevant variations with metallicity shown by the fitted data, but also be as small as possible to avoid degeneracies and computational overhead.

From the \argos data, \citet{Ness2013a} showed that the MDF and its field-to-field variations across the bar region could be well represented by only five numbers, representing the varying amplitudes of five Gaussian MDFs with approximately fixed means and dispersions. Hence those Gaussian MDFs, denoted $\alpha$, $\beta$, $\gamma$, $\delta$ and $\epsilon$ in the order of decreasing metallicity form a suitable set of EMDFs for modelling the full MDF of any particle. Note that we do not need to assume that the Gaussian EMDFs of \citet{Ness2013a} are the MDFs of actual physical components of the Galaxy with different origin and history. Their use is purely restricted to a mathematical set of EMDFs that allow modelling reasonable particle MDFs with only a few parameters per particle, as demonstrated later in \autoref{fig:ComponentsVsBins}.

For efficiency we neglect the Gaussian $\epsilon$ since it represents only about half a percent of all stars in the \argos survey and is thus very poorly constrained by the data. Note that the Gaussian EMDFs $\alpha$-$\delta$ were originally named A-D in \citet{Ness2013a}. We chose here a different naming convention to highlight the difference between the Gaussian EMDFs $\alpha$-$\delta$, and the metallicity bins A-D introduced by \citet{Ness2013b} and used later in \autoref{section:data}.

The MDF of each particle is then modelled as
\begin{equation}
 \label{equation:particleMDF}
 f_{\Feeq, i} = \sum_{c\in{\alpha - \delta}} w_{i,c} \mathcal{G}_{[\Feeq^0_c, \sigma_c]}
\end{equation}
where $w_{i,c}$ are the chemical weights, i.e. weights of the Gaussian EMDFs $c$ in the MDF of particle $i$, $\Feeq^0_c$ and $\sigma_c$ are the mean and standard deviation of the EMDFs and $\mathcal{G}_{[\mu, \sigma]}$ is the Gaussian distribution of mean $\mu$ and standard deviation $\sigma$. Adopted values for $\Feeq^0_c$ and $\sigma_c$ are quoted in \autoref{table:modelsParameters} and correspond to the means of the values determined by \citet{Ness2013a} (see their table 3).
\begin{table}
\caption{Parameters of the four Gaussian EMDFs used to describe the particle MDFs in our chemodynamical M2M modelling, based on table 3 of \citet{Ness2013a}.}
\label{table:modelsParameters}
  \centering
  \begin{tabular}{l|ccccc}
      & $\alpha$ & $\beta$ & $\gamma$ & $\delta$ \\
    \hline\hline
    $\Feeq^0_c$ & 0.11 & -0.28 & -0.69 & -1.18 \\
    $\sigma_c$ & 0.12 & 0.13 & 0.15 & 0.14 \\
  \end{tabular} 
\end{table}

From the MDF of every particle, we can evaluate any metallicity dependent observable $y_j$ on the model by
\begin{equation}
\label{equation:observableWithMetallicity}
 y_j = \sum_{i} w_i \times \sum_{c\in \alpha-\delta} K_{j,c}(\vec{x}_i,\vec{v}_i) w_{i,c}
\end{equation}
where 
\begin{equation}
\label{equation:finalObservableKernel}
 K_{j,c}(\vec{x}_i,\vec{v}_i) =  \int_{\Feeq} K_j(\vec{x}_i,\vec{v}_i, \Feeq) \, \times \mathcal{G}_{[\Feeq^0_c, \sigma_c]}(\Feeq) \,d\Feeq.
\end{equation}
The purpose of the new kernel function $K_j(\vec{x}_i,\vec{v}_i, \Feeq)$ is to limit the observable $y_j$ to a given volume in phase-space and apply on the model any observational bias that can exist on the real data which the model is compared to. We describe in detail how we model these observational biases for the \argos and \apogee surveys in \autoref{section:data}.

\subsection{M2M fit of the MDF}
\label{section:M2MfitMDF}
The M2M adaptation of the particle chemical weights $w_{i,c}$ is done in a similar fashion as the M2M evolution of the physical weights of the particles. The chemical weights $w_{i,c}$ are evolved in the direction of the gradient of the profit function $F$ according to
\begin{equation}
  \label{equation:FOCcomponents}
  \frac{dw_{i,c}}{dt} = \varepsilon_c \frac{\partial F}{\partial w_{i,c}}
\end{equation}
where $\varepsilon_c$ set the strength of the MDF fitting. 

As a profit function we adopt a simple chi-square given by 
\begin{equation}
 F = - \frac{1}{2} \sum_{j} \left ( \frac{y_j - Y_j}{\sigma(Y_j)} \right )^2
 \label{equation:profit}
\end{equation}
where $y_j$, $Y_j$ and $\sigma(Y_j)$, defined in more detail in \autoref{section:data}, are respectively the model observables, the data and its associated errors. The $w_{i,c}$ are additionally normalized to $\sum_c w_{i,c}  = 1$ at every step for every particle. Note that the weight adaptation described in \autoref{equation:FOCcomponents} requires some initial values for the chemical weights $w_{i,c}$ to start with, as discussed in more detail in \autoref{section:dynamicalModel}. For simplicity, we keep the particle masses $w_i$ constant throughout the M2M fit and only adapt the chemical weights $w_{i,c}$. This is justified since our initial model is already a good dynamical model of the entire galactic bar region, as described in the next section.

The M2M fit is performed in three successive phases. First, the model observables are initialized by evolving the N-body model for a time $\rm{T_{smooth}}$ and computing the temporally smoothed model observables. Then the model is evolved while adapting the chemical weights $w_{i,c}$ for a time $\rm{T_{M2M}}$. Finally the model is relaxed for a time $\rm{T_{relax}}$, to test the stability of the M2M fit. By using an internal time unit (iu) in which a circular orbit at $4\kpc$ has an orbital period of unity, we adopt for $\rm{T_{smooth}}$, $\rm{T_{M2M}}$ and $\rm{T_{relax}}$ the respective values of 4, 40 and 16 iu. The temporal smoothing time-scale is set to $\alpha = 1\,  \rm{iu}^{-1}$ and the strength of the MDF fitting to $\varepsilon_c = 10 \, \rm{iu}^{-1}$.

The weight evolution described by \autoref{equation:FOCcomponents} is performed iteratively by a first-order integration with small time step $\delta t = 10^{-3} \, \rm{iu}$. Between two iterations of the weight adaptation procedure, the N-body model is integrated forward using an adaptive drift-kick-drift leap-frog algorithm. The gravitational potential is computed directly from the particle distribution (stellar and dark matter particles) using the hybrid grid method of \citet{Sellwood1997}, modified in \hyperlink{P17}{P17} for allowing better vertical resolution in the bar region. We assume that the gravitational potential rotates at a constant pattern speed of $40\kmskpc$ corresponding to the pattern speed of the bar in the model of \hyperlink{P17}{P17} that matches both the \brava data in the bulge and the \argos data in the bar region. 

\section{Dynamical model and data constraints on the MDF}
\label{section:data}

In this section, we describe in detail the initial dynamical model used to provide the N-body orbits on to which we fit the particle MDFs in \autoref{section:dynamicalModel}, together with the two spectroscopic data sets, the \argos survey in \autoref{section:argos} and \apogee surveys in \autoref{section:apogee}. We then describe and compare our fiducial model to the data, and present variational models used to quantify systematic uncertainties.

\subsection{Dynamical model of the galactic bar region}
\label{section:dynamicalModel}
In \hyperlink{P17}{P17}, we presented the first non-parametric dynamical model of the entire bar region. This model reproduces the 3D density of RCGs in the bulge from \citet{Wegg2013}, the magnitude distributions of RCGs across the bar region from combination of the \vvv, \ukidss and \twomass surveys \citep{Wegg2015}, together with stellar kinematics from the \brava, \ogle and \argos surveys. Including the kinematics as a function of distance from the \argos survey we were able to recover the bar pattern speed and the mass distribution of both stars and dark matter in the inner $5\kpc$ of the Galaxy, hence constraining the full effective potential in the galactic bar region. We base our work on this model that provides a gravitational potential and a self-consistent library of N-body orbits. Note that throughout the chemodynamical M2M fit the particles masses and the gravitational potential in the bar frame are both kept fixed. 
\begin{figure}
  \centering
  \includegraphics{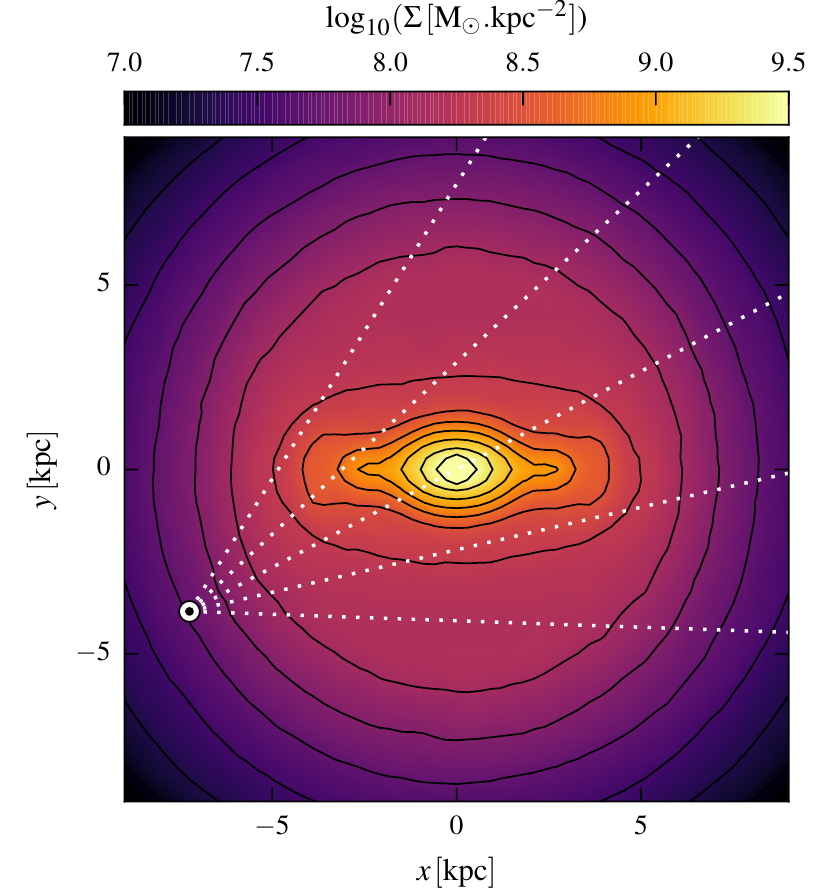}\\
  \caption{Face-on projection of the best-fitting model of \citet{Portail2016a}. The bar is at an angle of $\alpha=28\degree$ from the line of sight towards the Galactic Centre and rotates at $\Omega = 40 \kmskpc$. The dotted white lines originating from the Sun (dot symbol) indicate sightlines with galactic longitudes of $l=-30\degree, -15\degree, 0\degree, +15\degree$ and $+30\degree$.}
  \label{fig:SurfaceDensityBestModel}
\end{figure}
\autoref{fig:SurfaceDensityBestModel} shows the face-on surface density of this model. The bar rotates at $\Omega = 40\kmskpc$, as required to fit both the \brava kinematics in the bulge and the \argos radial velocity field in the bar region, and has in the model a half-length of $5.30 \pm 0.36\kpc$, in good agreement with the determination of bar the half-length in the Milky Way of $5.0\pm0.2\kpc$ by \citet{Wegg2015}. 

We introduce the metallicity of stars in the dynamical model by specifying for each particle an initial value for the four chemical weights $w_{i,c}$. We call these initial weights the ``prior'' MDF although they do not correspond to a formal Bayesian prior. It is a priori unclear how much the final result of a chemodynamical M2M fitting depends on the assumed prior MDF, as the chemical weights $w_{i,c}$ are only allowed to change when travelling through a portion of space constrained by the data. Since we focus on the bar region that is well covered by the \argos survey, our fiducial choice for the prior MDF is the fractions that correspond to the overall MDF of all the \argos stars: $w_{i,\alpha} = 0.23$, $w_{i,\beta} = 0.43$, $w_{i,\gamma} = 0.29$ and $w_{i,\delta} = 0.05$. We consider alternative priors and quantify their effects on the final results in \autoref{section:fiducialAndVariations}.

The bar is oriented at an angle of $28\degree$ to the line of sight towards the Galactic Centre \citep{Wegg2015}. Following the recommendations of \citet{BlandHawthorn2016}, we place the Sun at $R_0 = 8.2\kpc$ and assume that the Local Standard at Rest (LSR) is on a circular orbit at $V(R_0) = 238\kms$. We adopt for the peculiar motion of the Sun in the LSR the values of $(U,V,W) = (11.1, 12.24, 7.25) \kms$ as measured by \citet{Schonrich2010}. Altogether these assumptions lead to a proper motion of the Galactic Centre of $30.5\kmskpc$ consistent with the total angular velocity on the Sun of $30.57\pm0.43\kmskpc$ measured from proper motion of masers by \citet{Reid2014}. All radial velocities quoted in this work are expressed in the Galactocentric inertial frame. We re-transformed the radial velocity measurements of the \argos and \apogee surveys using our set of assumptions since they differ from the traditional transformations used by many spectroscopic surveys. In order to predict the selection function of the surveys and apply them when observing the model, we use the PARSEC isochrones \citep{Bressan2012, Chen2014,Tang2014} and assume a $10\Gyr$ old population for  a Kroupa IMF.

\subsection{\argos as a  function of distance and metallicity}
\label{section:argos}

\begin{figure*}
  \centering
  \includegraphics{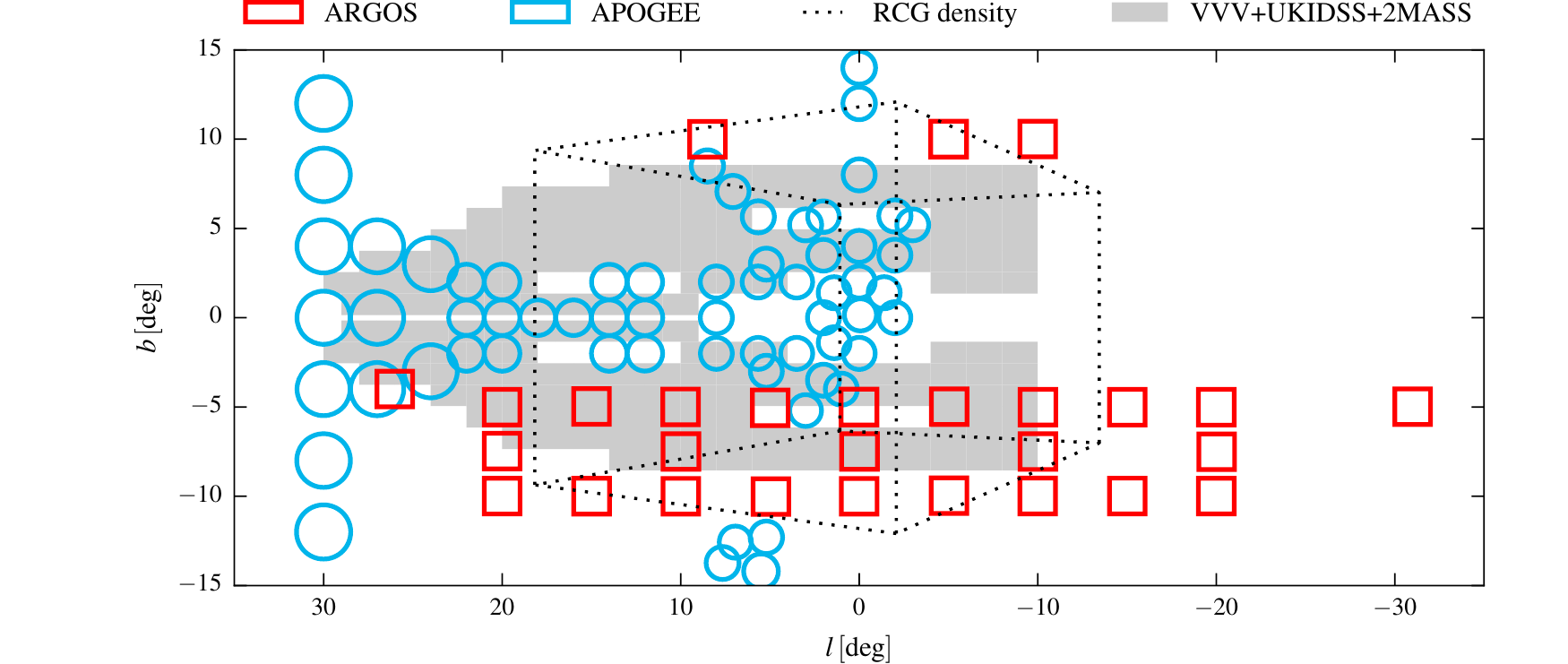}\\
  \caption[]{Spatial coverage in galactic coordinates of the data sets used to constrain the MDF in the model. Red squares show the \argos fields, constraining the three-dimensional spatial and kinematic variations of metallicity (i.e. on the sky and also as a function of line of sight distance, see \autoref{section:argos}). Blue circles show the \apogee (DR12) fields, constraining the spatial and kinematic variations of metallicity integrated along the line-of-sight (\autoref{section:apogee}). The dashed box and grey regions show the coverage of the 3D density of RCGs in the bulge and the magnitude distributions of RCGs from \vvv + \ukidss + \twomass used in \hyperlink{P17}{P17} to constrain the dynamical model.}
  \label{fig:fieldsPosition}
\end{figure*}

The Abundance and Radial velocity Galactic Origin Survey (\argos) is a large spectroscopic survey of about $28000$ stars towards the galactic bulge and inner disk, designed to sample RCGs all the way from the nearby disk ($\sim 4.5\kpc$ from the Sun) to the far side of the Galactic Centre ($\sim 13\kpc$ from the Sun). The survey covers lines of sight at longitudes between $l=-20\degree$ and $l=20\degree$ and latitudes between $b=-5\degree$ and $-10\degree$, as well as a few northern fields, as shown in \autoref{fig:fieldsPosition}.
The \argos stars are selected from the \twomass catalogue according to a procedure fully described in \citet{Freeman2013} for which we modelled the selection function in \hyperlink{P17}{P17}. In brief, the \argos selection function consists of (i) a weight for each star and (ii) a function of the extinction corrected magnitude $C({K_{s}}_0)$ for each field. The set of weights comes from the fact that for each field an equal number of stars were selected in three magnitude bins in order to sample the full range of distance from the nearby disk to the far side of the bulge. The selection function as a function of magnitude represents the probability for a star to pass the various selection criteria of the survey, thus taking into account the magnitudes and colour cuts, high-quality imaging criteria and incompleteness of the input catalogue that altogether bias the survey towards nearby stars. Details about the modelling of the \argos selection function can be found in \hyperlink{P17}{P17}.

From the medium resolution spectra of the \argos stars, \citet{Ness2013a} estimated various stellar parameters including the radial velocity and the metallicity with a typical error of $1\kms$ and $0.1 \, \rm{dex}$ respectively. Since the \argos stars are primarily RCGs which are approximate standard candles, we can infer accurate stellar distances from their photometry. We adopt an absolute magnitude of $M_{K_s, \rm{RCG}} = -1.72$ as used by \citet{Wegg2013} for the bulge, and bin stars as a function of distance. In \hyperlink{P17}{P17}, we constructed the \argos radial velocity field by computing for each survey field the mean velocity and velocity dispersion of the stars in several distance modulus bins of $0.25 \, \rm{mag}$ along the line of sight. We adopt here the same distance bins as \hyperlink{P17}{P17}.

Since the number of stars per distance bin is typically $\sim 80-100$ in each field, we can split the bins further in the metallicity space. Following \citet{Ness2013b}, we define four metallicity bins of $\Delta \Fe = 0.5\, \dex$ called A-D by $0.5 \geq \Fe > 0$ (A), $ 0 \geq \Fe > -0.5$ (B), $-0.5\geq \Fe > -1$ (C) and $-1 \geq \Fe > -1.5$ (D). Binning the stars avoids correlations between data points and is therefore well suited to the $\chi^2$ minimization of the M2M method.

\begin{figure}
  \centering
  \includegraphics{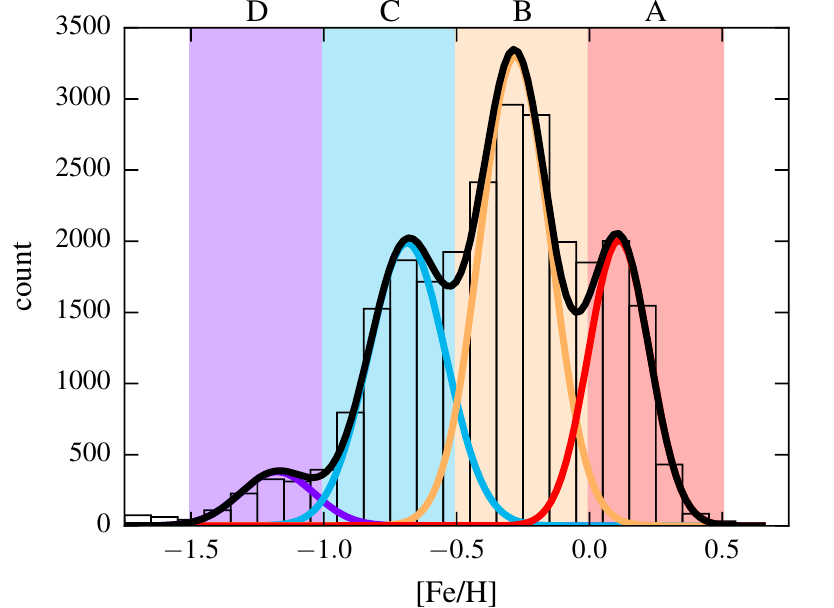}\\
  \caption{Gaussian EMDFs $\alpha-\delta$ versus metallicity bins A-D. The observed example MDF (histogram) is binned in four bins A-D shown by the colour bands. From the number counts in all bins, we can compute the four chemical weights $w_c$ of the Gaussians EMDFs $\alpha-\delta$ (coloured solid lines, see \autoref{table:modelsParameters}) for which the integral of the total MDF has the same count as the stars in each metallicity bin. The resulting MDF (black line) is a very good representation of the underlying histogram and justifies our use of the Gaussians $\alpha-\delta$ as a set of EMDFs for modelling the particle MDFs.}
  \label{fig:ComponentsVsBins}
\end{figure}

The difference between the Gaussian EMDFs $\alpha-\delta$ introduced in \autoref{section:particleMDF} and the metallicity bins A-D introduced here is best described by \autoref{fig:ComponentsVsBins}. In this figure, the histogram shows a typical MDF observed in a field towards the bulge. By binning the stars in the four bins A-D, we obtain a low resolution version of the MDF. However, this low resolution MDF is not a faithful representation of the real underlying MDF, which for example in bin A has many more stars close to the lower end of the bin than stars close to the higher end of the bin. A more faithful representation of the MDF is obtained using the set of EMDFs $\alpha-\delta$. Indeed, from the counts in the four bins A-D, we can determine four unique numbers $w_c$, corresponding to the weights of the Gaussian EMDFs $\alpha$-$\delta$, for which the integral of the total MDF has the same counts as the stars in each of the metallicity bins. Summing up the Gaussian EMDFs, we obtain a full MDF, using only these four numbers, that is a very good representation of the underlying histogram as shown in \autoref{fig:ComponentsVsBins}. The relation between the counts in bins A-D and the chemical weights of \autoref{section:particleMDF} is given by a $4 \times 4$ matrix, whose elements correspond to the contribution of each Gaussian EMDFs to each bin. Given our choices of bins and Gaussian EMDFs, this matrix is mostly diagonal.

For a field indexed by $j$, distance bin indexed by $d$ and metallicity bin indexed by $m$ we compute the following three quantities:
\begin{enumerate}
 \item The fraction $f_{j,d,m}$, of the observed stars in distance bin $d$ of field $j$ that have metallicities within bin $m$
 \item Mean velocity $v_{j,d,m}$ of the stars considered above
 \item Velocity dispersion $\sigma_{j,d,m}$ of the stars considered above
\end{enumerate}

Errors in $f_{j,d,m}$, $v_{j,d,m}$ and $\sigma_{j,d,m}$ are estimated respectively as $1/\sqrt{n}$, $\sigma_{j,d,m}/\sqrt{n}$ and $\sigma_{j,d,m}/\sqrt{2 \, n}$ for $n$ the number of stars in bin $(j,d,m)$. Since $\sigma_{j,d,m}$ is biased towards lower values for small number statistics, we exclude from further considerations the kinematics of bins with less than five stars. In some fields and metallicity bins, there are not enough stars in the \argos sample to also bin in distance. Hence for each field $j$ and metallicity bin $m$, we compute in addition the three similar quantities $f_{j,m}$, $v_{j,m}$ and $\sigma_{j,m}$ corresponding to the fraction, mean velocity and velocity dispersion of stars integrated along the line of sight. $f_{j,d,m}$, $v_{j,d,m}$ and $\sigma_{j,d,m}$ are shown together with our fiducial chemodynamical model described in \autoref{section:fiducialAndVariations}. 

\begin{figure}
  \centering
  \includegraphics{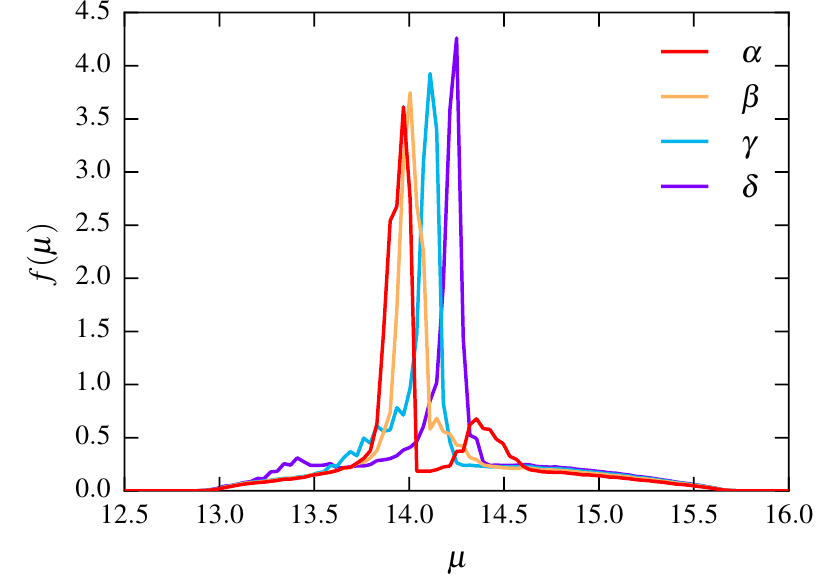}\\
  \caption{Distance distribution of observable mock stars drawn from a particle located at distance modulus $\mu = 14.0$, for the metallicity distributions of the Gaussian EMDFs $\alpha-\delta$ after applying the \argos selection function. Metal-rich RCGs are typically brighter in the near-infrared than more metal-poor stars RCGs and appear therefore closer. The secondary peak in $f(\mu)$ for EMDF $\alpha$ is due to the red giant branch bump which is prominent for the most metal-rich population \citep[e.g.][]{Nataf2011, Nataf2013a}.}
  \label{fig:distanceDistributionOfComponents}
\end{figure}

In order to compare the model to the data, we need to form model observables that are the model equivalent quantities of $f_{j,d,m}$, $v_{j,d,m}$ and $\sigma_{j,d,m}$. To do so, we first turn the model particles into mock stars, then apply the \argos selection function and finally estimate the distance of the remaining mock stars assuming they are RCGs, as done for the observed stars. This is described in detail in \hyperlink{P17}{P17} and will only be outlined here. For a given particle, we simulate a mock $10\Gyr$ old stellar population for a Kroupa IMF and use the PARSEC isochrones in combination with the selection function to predict the magnitude distribution of \argos mock stars, i.e. RCGs on top of a background of giant stars. We then apply the distance estimation based on the RCG magnitude as we do for the \argos stars and compute the distance distribution of observable mock stars drawn from the particle. This distribution, called $f_i(\mu)$ in \hyperlink{P17}{P17}, encloses all selection and stellar population effects but does not depend on metallicity. Here, we extend $f_i(\mu)$ to $f_i(\mu, \Feeq)$ taking into account the fact that the luminosity function of different metallicities is slightly different: more metal-rich RCGs are brighter in the near-infrared than more metal-poor RCGs. The effect of metallicity is illustrated in \autoref{fig:distanceDistributionOfComponents} where we show the distance distributions of observable mock stars corresponding to a particle located at $\mu = 14.0$ for different metallicity distributions corresponding to the four Gaussian EMDFs $\alpha-\delta$. More metal-rich stars appear at closer distances than more metal-poor stars.

 Motivated by the general form of an observable given in \autoref{equation:observableWithMetallicity}, we finally define our \nmagic observables as the number count and first and second mass-weighted velocity moments. The \nmagic kernel of the observables to be used in \autoref{equation:finalObservableKernel} are then given by the following equations for a field $j$, a distance bin $d$ and a metallicity bin $m$:
\begin{equation}
\label{equation:kernelArgosNumber}
 K_{j, d, m}^{\rm{\argos}, 0}(\vec{z}_i) = \delta_{j, d, m}^{\rm{\argos}}(\vec{z}_i)
\end{equation}
\begin{equation}
K_{j, d, m}^{\rm{\argos}, 1}(\vec{z}_i) = \frac{ \delta_{j, d, m}^{\rm{\argos}}(\vec{z}_i) }{W_{j, d, m}^{\rm{\argos}}} \times {v}_i
\end{equation}
and 
\begin{equation}
K_{j, d, m}^{\rm{\argos}, 2}(\vec{z}_i) = \frac{ \delta_{j, d, m}^{\rm{\argos}}(\vec{z}_i)}{W_{j, d, m}^{\rm{\argos}}} \times {v}_i^2
\end{equation}
where $\vec{z}_i = (\vec{x}_i, \vec{v}_i, \vec{M}_i)$ is the extended phase-space vector of particle $i$ and ${v}_i$ its radial velocity. $W_{j,d, m}^{\rm{\argos}}$ is given by 
\begin{equation}
W_{j,d, m}^{\rm{\argos}} = \sum_i w_i \delta_{j,d,m}^{\rm{\argos}}(\vec{z}_i)
\end{equation}
 and $\delta_{j,d,m}^{\rm{\argos}}(\vec{z}_i)$ by
\begin{equation}
\label{equation:deltaArgos}
 \delta_{j,d,m}^{\rm{\argos}} = 
 \begin{cases}
   \int_{\mu(d)}^{\mu(d+1)} f_i(\mu, \Feeq)\, d\mu & \text{if $i \in $ field $j$ }\\
    & \text{and $\Feeq$ in bin $m$,}\\
   0 & \text{otherwise}
  \end{cases}
\end{equation}
with $\mu(d)$ and $\mu(d+1)$ the boundaries of the distance modulus bin $d$. \nmagic kernels for data integrated along the line of sight are defined in a similar way.

These kernels are then inserted into equations (\ref{equation:observableWithMetallicity}) and (\ref{equation:finalObservableKernel}) to give the full expression for a model observable. In the case of the model observable `number count in field $j$, distance bin $d$ and a metallicity bin $m$', denoted $y_{j,d,m}^{\rm{\argos}, 0}$, this insertion sums up to
\begin{equation}
\begin{split}
 &y_{j,d,m}^{\rm{\argos}, 0} = \sum_{i} w_i \times \sum_{c\in \alpha-\delta} w_{i,c} \times \\
& \int_{\Feeq \in m} \int_{\mu(d)}^{\mu(d+1)} f_i(\mu, \Feeq) \times \mathcal{G}_{[\Feeq^0_c, \sigma_c]}(\Feeq) \, d\mu d\Feeq
\end{split}
\end{equation}
and similarly for the first and second mass weighted velocity moments.

\subsection{\apogee as a function of metallicity}
\label{section:apogee}
The Apache Point Observatory Galactic Evolution Experiment (\apogee) is a high resolution spectroscopic survey designed to sample a large number of stars ($\sim 1.5\times 10^5$) in all possible galactic environments, from the stellar halo to the disk and the bulge \citep{Eisenstein2011, Majewski2012, Nidever2012}. \apogee operates in the near-infrared and is therefore able to pierce through highly dust obscured regions, making it particularly useful close to the galactic plane. Among the $\sim 1.5 \times 10^5$ stars about $10^4$ are located in fields towards the bulge and long bar, some of which are directly in the galactic plane. We thus use \apogee to complement \argos in the galactic plane, in the fields shown in \autoref{fig:fieldsPosition}. We use the stellar parameters and spectroscopic distances of stars in the \apogee bulge and bar fields determined by \citet{Ness2016} from the publicly available spectra of DR12 \citep{GarciaPerez2013, Holtzman2015, Shetrone2015} using the data-driven method \emph{The Cannon} \citep{Ness2015}. The \apogee bar and bulge fields do not contain enough stars for binning both in distance and metallicity. We thus consider for each field only a single distance bin ranging from $4\kpc$ to $12\kpc$ from the Sun and bin the stars in the four metallicity bins A-D. As in \autoref{section:argos}, we construct for each \apogee field the quantities $f_{j,m}$, $v_{j,m}$ and $\sigma_{j,m}$ corresponding respectively to the fraction of observed stars of field $j$ that have metallicities within bin $m$, their mean velocity and their velocity dispersion.

The \apogee survey is targeted at bright red giants selected from the \twomass catalogue according to a complex procedure fully described in \citet{Zasowski2013}. To fairly compare the model to the \apogee data, we need to model the distance and metallicity biases introduced by the \apogee selection criteria. Luckily \citet{Bovy2016} and \citet{Bovy2016a} computed and made available the \apogee selection function for any kind of tracer stellar population. Using the PARSEC isoschrones to simulate a mock $10\Gyr$ old stellar population with a Kroupa IMF at a given metallicity  $\Feeq$, the \apogee selection function provides us with the fraction of stars observable by the \apogee survey as a function of distance $\mu$. This fraction corresponds to the quantity $f_i(\mu, \Feeq)$ already used to define the \argos kernels in \autoref{equation:deltaArgos}. Hence, our \apogee kernels are basically the same as \autoref{equation:deltaArgos} with the exception that we have only one large bin in distance covering all the way from $4\kpc$ to $12\kpc$ along the line of sight. $f_{j,m}$, $v_{j,m}$ and $\sigma_{j,m}$ are shown together with our fiducial M2M chemodynamical model in the next subsection.

\subsection{Fiducial and variation models}
\label{section:fiducialAndVariations}

\begin{figure}
  \centering
  \includegraphics{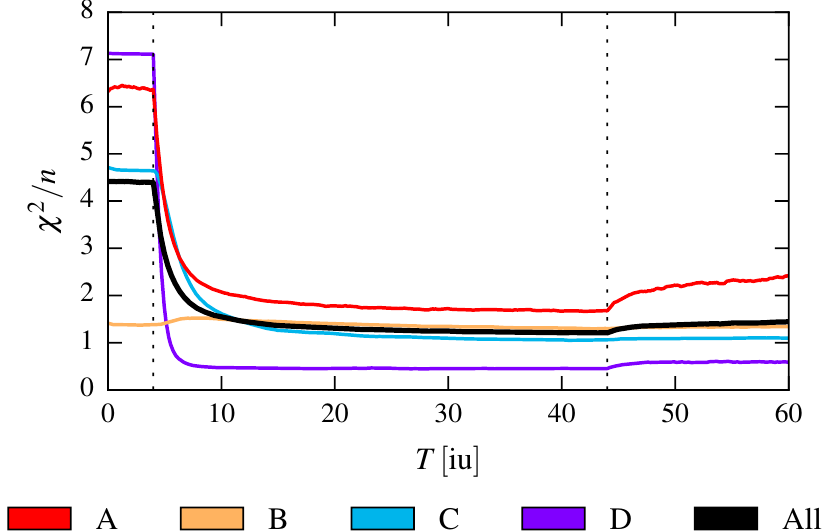}\\
  \caption{Evolution of the $\chi^2$ during the M2M fit for observables in the four metallicity bins separately (coloured lines) and all observables together (black line). The black dotted lines mark respectively the end of the initial smoothing phase at $T = 4\, \rm{iu}$ and the beginning of the relaxation phase at $T = 44\, \rm{iu}$.}
  \label{fig:chi2evolution}
\end{figure}

Our fiducial model is obtained by fitting the best-fitting dynamical model of \hyperlink{P17}{P17} to the \argos and \apogee (DR12) data using the M2M chemodynamical method described in \autoref{section:chemodynamicalModelling}. \argos and \apogee have a different spatial coverage and complement each other in constraining the entire galactic bar region, as shown by \autoref{fig:fieldsPosition}. The evolution of $\chi^2$ during the M2M fit is shown in \autoref{fig:chi2evolution}, for observables in the four metallicity bins separately. It shows the typical evolution of $\chi^2$ in M2M modelling, staying constant during $\rm{T_{smooth}}$, decreasing during $\rm{T_{M2M}}$ while fitting, to finally slightly increase again and stabilize during  $\rm{T_{relax}}$.

\begin{figure*}
  \centering
  \includegraphics{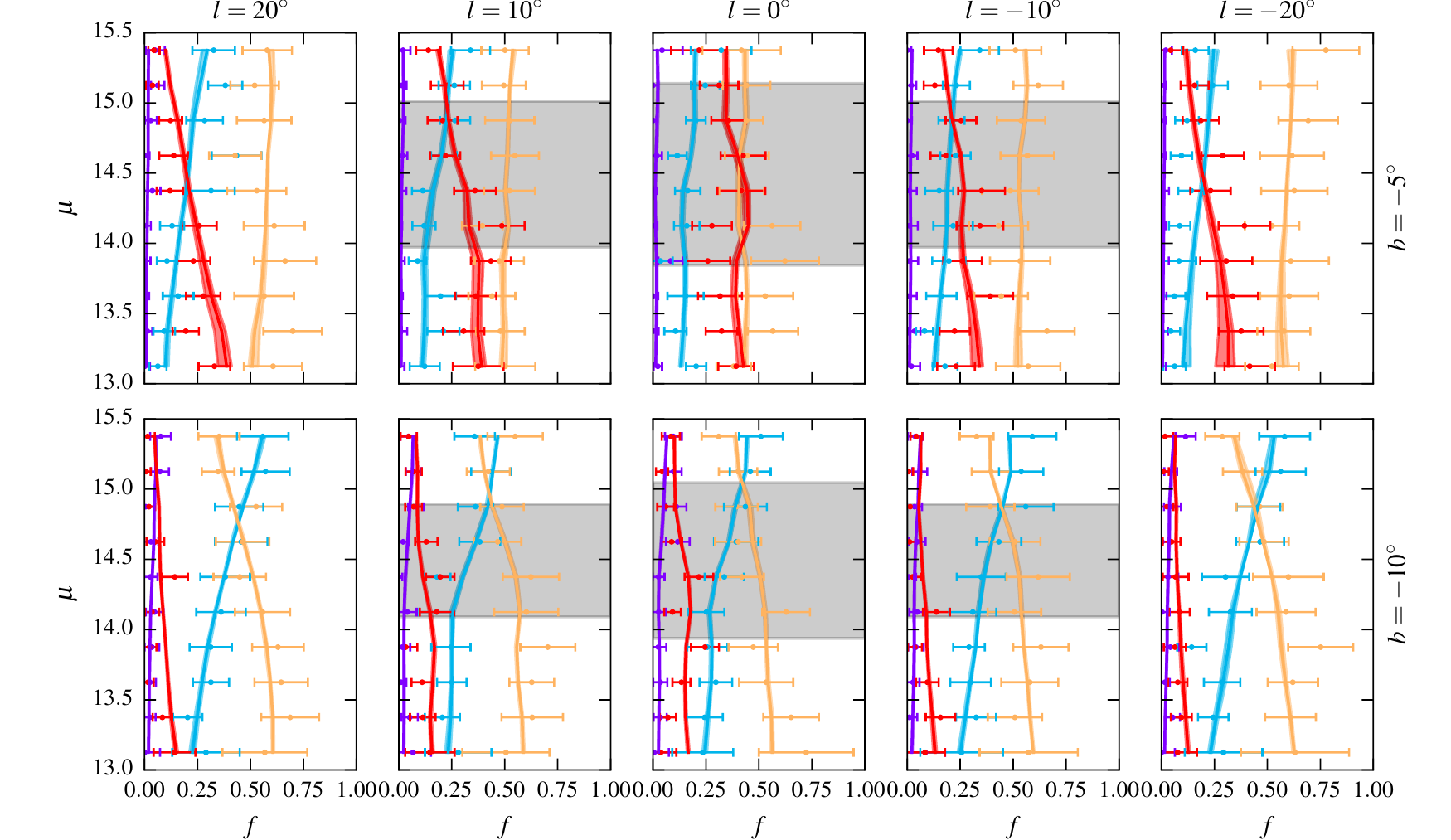}\\
  \medskip
  \includegraphics{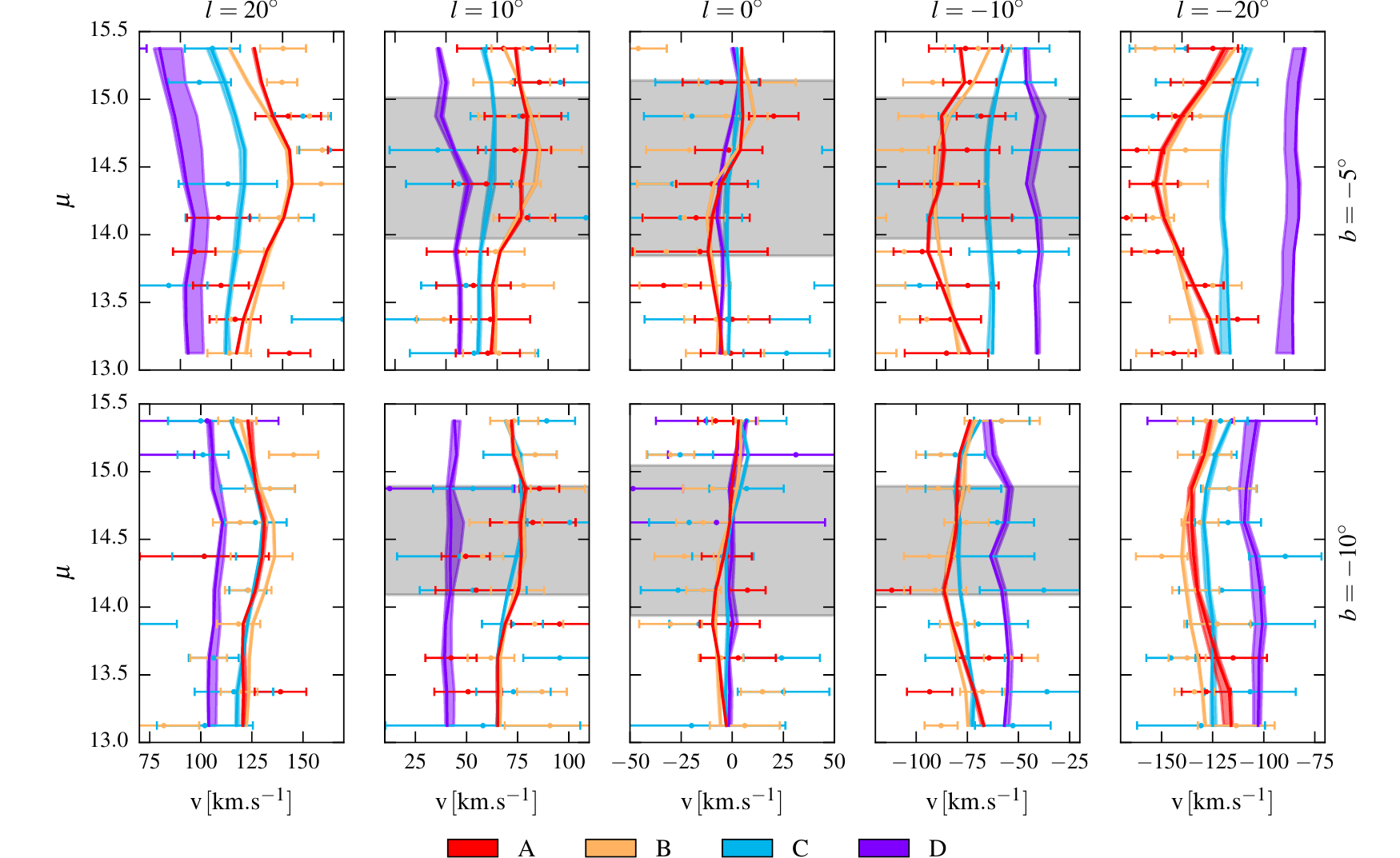}\\
  \caption{Panels from top to bottom show the fraction of stars $f$, mean velocity $v$ and velocity dispersion $\sigma$ in each metallicity bin as a function of distance modulus $\mu$ for our fiducial model (solid lines), compared to the \argos data (error bars). The first (second) row of each panel shows fields at $b=-5\degree$ ($b=-10\degree$) for $l$ between $20\degree$ and $-20\degree$, thus covering the bar region. The coloured areas indicate the range of values spanned by the five variations models. For the kinematics, only bins with more than five stars are shown, although the models are plotted everywhere. The grey area indicates the boundary of the volume within $2.5\kpc$ from the Galactic Centre.}
  \label{fig:ARGOS}
\end{figure*}

\begin{figure*}
  \centering
  \includegraphics{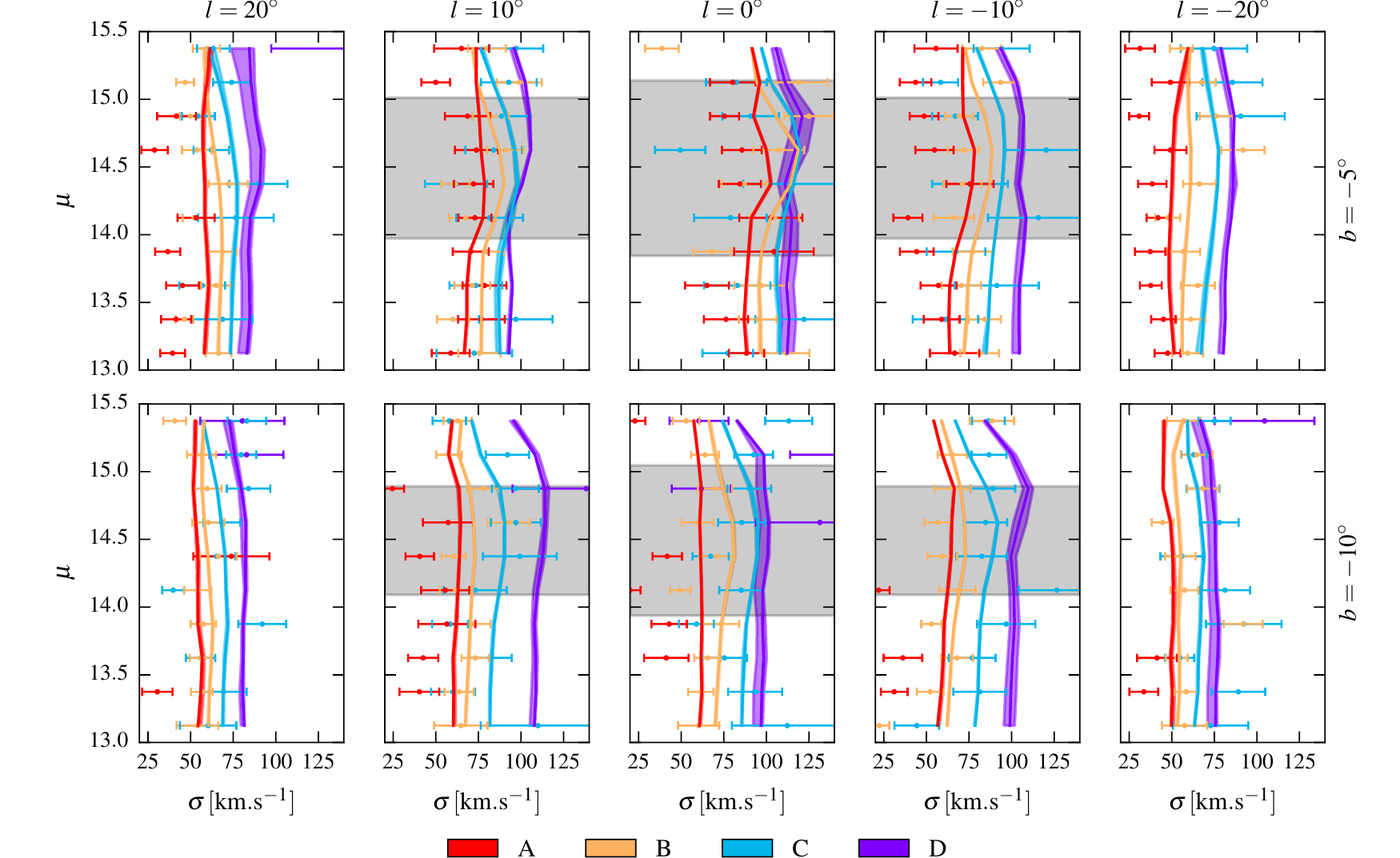}\\
  \contcaption{}
\end{figure*}

The three panels of \autoref{fig:ARGOS} show respectively the comparison between our fiducial model and the \argos data for the fraction of stars in each metallicity and distance bin $f_{j,d,m}$, their mean radial velocity $v_{j,d,m}$ and radial velocity dispersion $\sigma_{j,d,m}$. The distance distribution of the metallicity bins B and C are remarkably well fitted. Our model slightly overpredicts A in the central field along the minor axis at $b=-5\degree$, which is also the field most affected by selection effects due to the important extinction and crowding. The kinematics is also well fitted and reproduces the main trends as a function of metallicity already described by \citet{Ness2013b}: B is a hotter and slightly faster rotating replica of A while C and D are hotter and slower than both B and A at $|b| = 5\degree$. We note however that the model systematically overpredicts the dispersion of the very cold metallicity bin A by about $10\kms$.

\begin{figure*}
  \centering
  \includegraphics{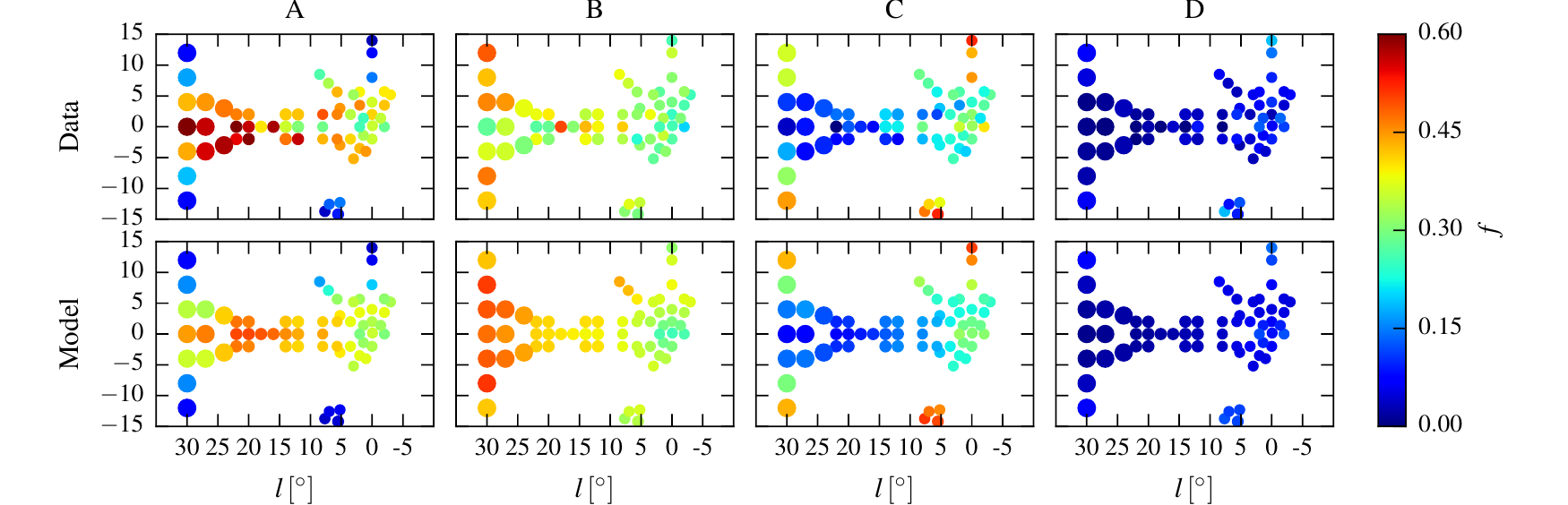} \\
  \bigskip
  \includegraphics{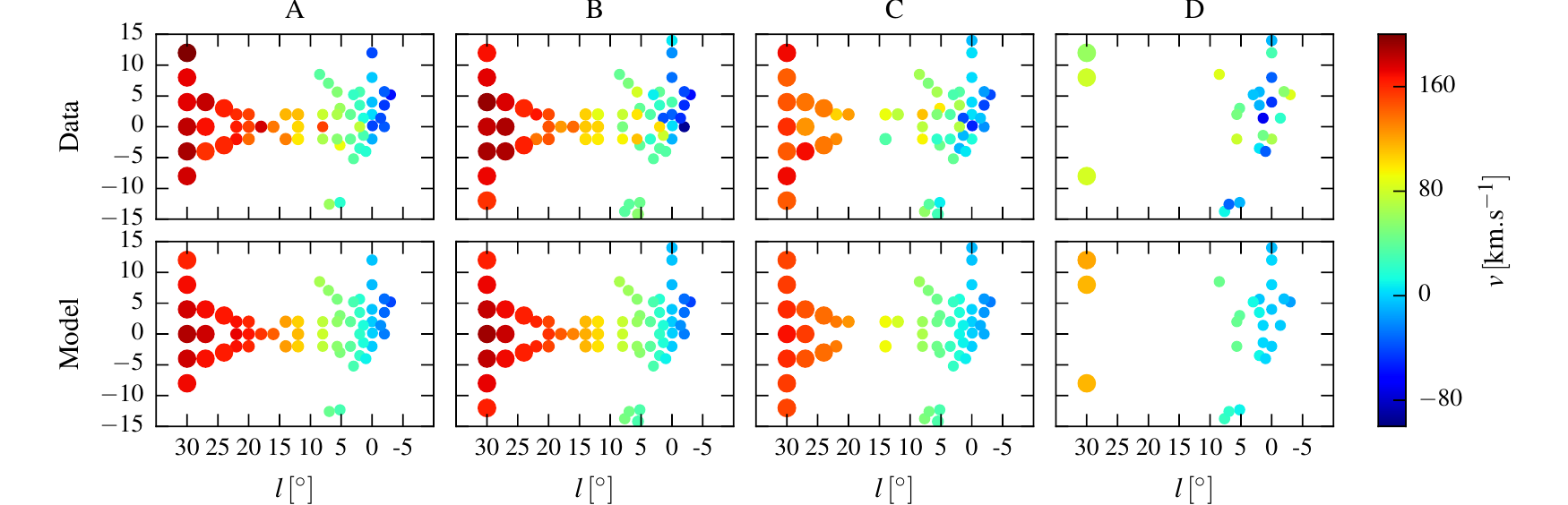}\\
  \bigskip
  \includegraphics{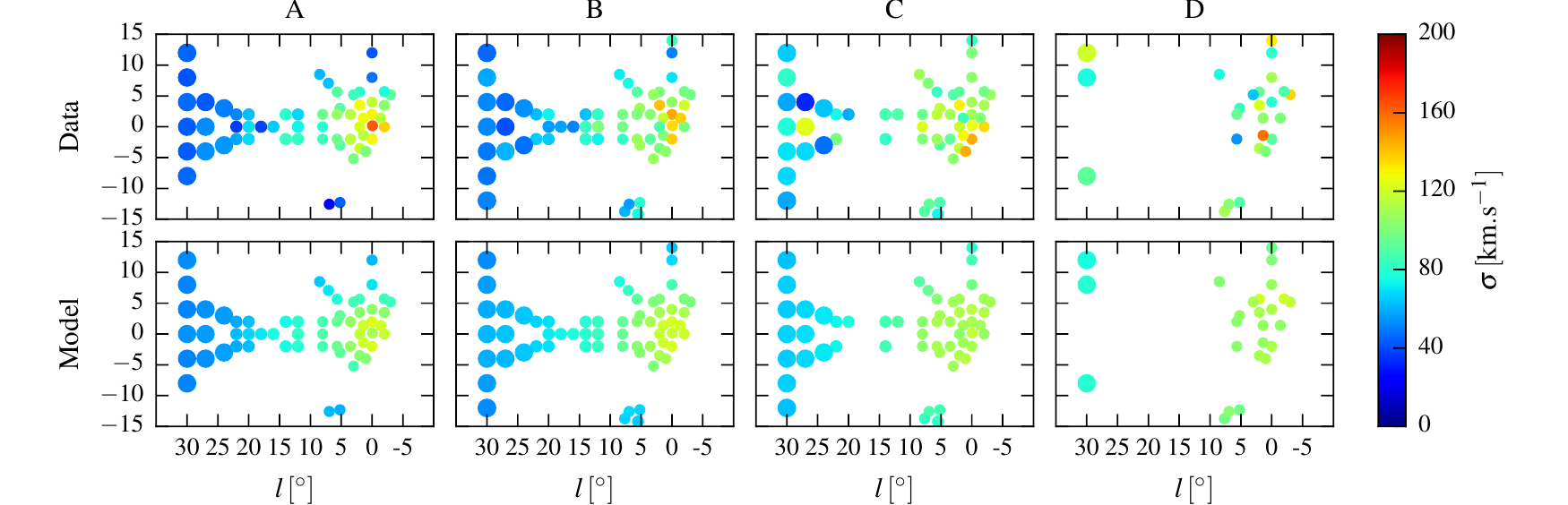}\\
  \caption{Panels from top to bottom show the fraction of stars, mean velocity and velocity dispersion in the four metallicity bins (columns) in the \apogee (DR12) data (first row in each panel) and in the model (second row in each panel).}
  \label{fig:APOGEE}
\end{figure*}

Similarly, the three panels of \autoref{fig:APOGEE} show respectively the comparison of our model with the fraction $f_{j,m}$, mean radial velocity $v_{j,m}$ and radial velocity dispersion $\sigma_{j,m}$ as computed from the \apogee data. In all metallicity bins, the model performs well at reproducing the data. We note however in metallicity bin A that the model tends to systematically underpredict the fraction of stars, an opposite trend to that found in the \argos data. This disagreement is probably due to a slight inconsistency between the two data sets, possibly originating from systematic effects in the selection function or from a difference in the upper end of the metallicity scales between \argos and \apogee.

Our fiducial model is based on modelling assumptions whose effects need to be quantified before analysing the spatial and kinematic distribution of stars in different metallicity bins. To this end we repeated the modelling for five alternative models with different assumptions as described below:
\begin{description}
 \item[(i) and (ii):] As already mentioned in \autoref{section:dynamicalModel}, part of the Galaxy is not constrained by our data sets and therefore the model partly depends on the prior MDF distribution assumed. We thus consider two variations of the prior distribution, the first corresponding to ``flat-priors'' where we initialize all $w_{i,c}$ to $0.25$ and a second more complex corresponding for each particle to the values determined from the MDF at the position of the particle in the Besan\c{c}on Galaxy Model \citep{Robin2003}.
 \item[(iii):] The \apogee sample is subject to a severe metallicity bias in the inner Galaxy, as discussed in the appendix of \citet{Hayden2015} and shown very clearly in fig. 2 of \citet{Ness2016}. Only a few stars in the \apogee fields towards the bulge are actually in the bulge, and those are preferentially metal-poor stars. This bias against metal-rich stars is already modelled by the \apogee selection function but is also somewhat uncertain since it relies on stellar population models and assumptions about the IMF and age of the stellar population in the inner part of the Galaxy. \citet{Hayden2015} showed that this bias could be reduced by removing from the sample all stars with $\log g\leq1\, \dex$. We thus repeat the fitting using this cut in surface gravity and modify the selection function accordingly.
 \item[(iv) and (v):] When using \apogee in combination with \argos we might expect a systematic difference in the metallicity scales of the two surveys. Such differences in scale are always present between different spectroscopic surveys and \citet{Smiljanic2014} showed that even when analysing the exact same spectra, different working groups would measure $\Fe$ with systematic variations of $\sim 0.1\, \dex$. Garc\'{i}a Perez et al. (in preparation) find a possible indication for a difference in metallicity scale between \argos and \apogee, based on about 100 stars in common between the two surveys. However, this comparison is limited to the brightest \argos stars in only one field and thus does not allow the precise characterization of the overall relation between the metallicity scales of the two surveys. Thus, we limit ourselves here to considering two alternative \apogee data sets where values for $\Fe$ are offsets by $\pm 0.1\, \dex$.
\end{description}
The difference between these variation models is found to be relatively small in the region covered by our data, as shown by the coloured areas in \autoref{fig:ARGOS}. In all the following, we will use the standard deviation of the results provided by the fiducial model plus its five variations as a measure of the systematic uncertainty affecting our results. \autoref{table:masses} gives an overview of all models and provides the stellar mass within $5\kpc$ from the Galactic Centre in all metallicity bins for the fiducial and variations models.

\begin{table}
\caption{Stellar masses within $5\kpc$ from the Galactic Centre for all models and metallicity bins, given in units of $10^{10}\, \Msun$}
\label{table:masses}
  \centering
  \begin{tabular}{l|cccc}
    Models & A & B & C & D\\
    \hline\hline
    Fiducial & 1.16 & 1.11 & 0.37 & 0.04\\
    Variation with flat priors & 1.20 & 1.07 & 0.37 & 0.04\\
    Variation with BGM priors & 1.10 & 1.16 & 0.39 & 0.04\\
    Variation with $\log g\leq1\, \dex$ & 1.12 & 1.14 & 0.39 & 0.04\\
    Variation with $\Fe$ offset of $+ 0.1\, \dex$ & 1.19 & 1.11 & 0.35 & 0.04\\
    Variation with $\Fe$ offset of $- 0.1\, \dex$ & 1.03 & 1.17 & 0.42 & 0.06\\
  \end{tabular} 
\end{table}

\section{Spatial distribution of the metallicity in the inner Galaxy}
\label{section:spatialVariations}
In this section, we show how the spatial distributions of stars in the Galaxy vary as a function of metallicity. Since all particles carry a full continuous MDF, we can in principle observe the model at any metallicity. To ease the comparison with other work, we present our model in the four metallicity bins A - D. To recall, the bins A, B, C and D correspond respectively to $0.5 \geq \Fe > 0$ (A), $ 0 \geq \Fe > -0.5$ (B), $-0.5\geq \Fe > -1$ (C) and $-1 \geq \Fe > -1.5$ (D).

\subsection{The 3D distribution of metallicity}
\label{section:spatialProjections}

\begin{figure*}
  \centering
  \includegraphics{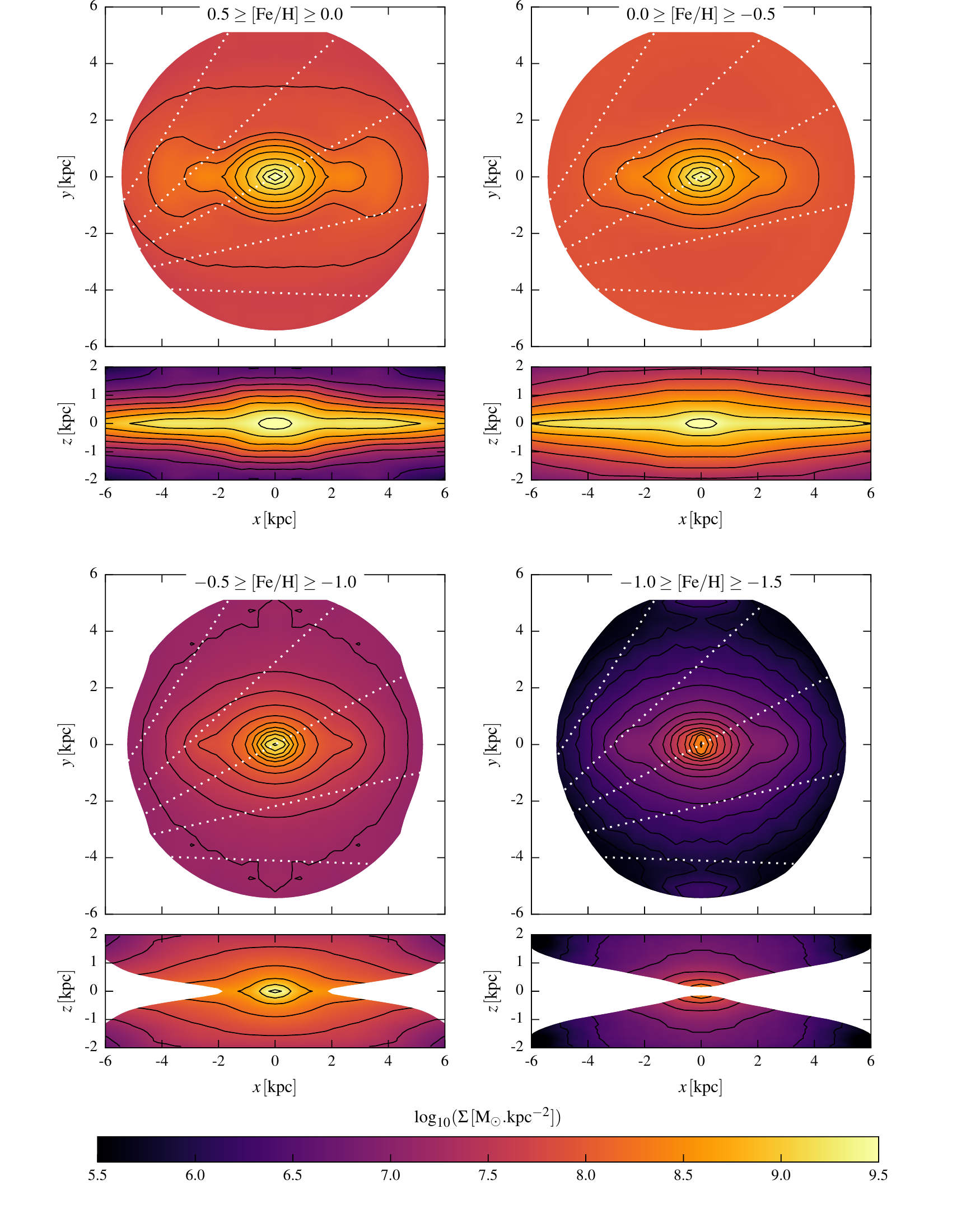}\\
  \caption{Face-on and side-on surface densities of the fiducial model in the four metallicity bins obtained after fitting the \argos and \apogee chemokinematic data. Regions where systematic uncertainties are larger than $30\%$ or outside $5.5\kpc$ from the Galactic Centre are masked. The white dotted lines indicate lines of sight at $l=30\degree$, $15\degree$, $0\degree$, $-15\degree$ and $-30\degree$.}
  \label{fig:allSurfaceDensitiesOfComponents}
\end{figure*}

The surface density in face-on and side-on projections of our fiducial model in the four metallicity bins A-D are shown in \autoref{fig:allSurfaceDensitiesOfComponents}. White regions indicate parts of the Galaxy where the systematic uncertainties are larger than $30\%$, or outside the central $5.5\kpc$. Bin A stars are concentrated to the plane and are strongly bar shaped. The hammer-like shape of the bar ends is similar to the shape of the superthin bar component, discovered in the \vvv + \ukidss + \twomass data by \citet{Wegg2015}. This hammer-like shape is confined to the plane and although being very clear in the mid-plane density, it is usually absent in the total surface density because of the mass distribution above and below the plane (see \autoref{fig:SurfaceDensityBestModel} for example). We thus speculate that the superthin bar component is significantly metal rich. Bin B stars are also strongly bar shaped but are less concentrated to the plane and do not significantly contribute to the superthin bar.

For metallicity bins C and D, the stellar densities are much more extended vertically, are both less bar shaped but do not show any significant B/P shape in the bulge. Outside the central $\kpc$, from its side-on projection in \autoref{fig:allSurfaceDensitiesOfComponents}, the stellar density in bin C is found to have a vertical flattening of $\sim 0.3$. Hence, at these radii, it has the shape of a thick disk. However in the central $\kpc$, stars in bins C and D appear to be consistently more centrally concentrated than the inward extrapolation of their outer exponential density profiles, as discussed in more detail in \autoref{section:centralConcentrationCandD}. Note also that only a small fraction of stars from our data sets belong to bin D (<5\% of the stars in \argos) and therefore our results for this metallicity bin are more uncertain.

\begin{figure*}
  \centering
  \includegraphics{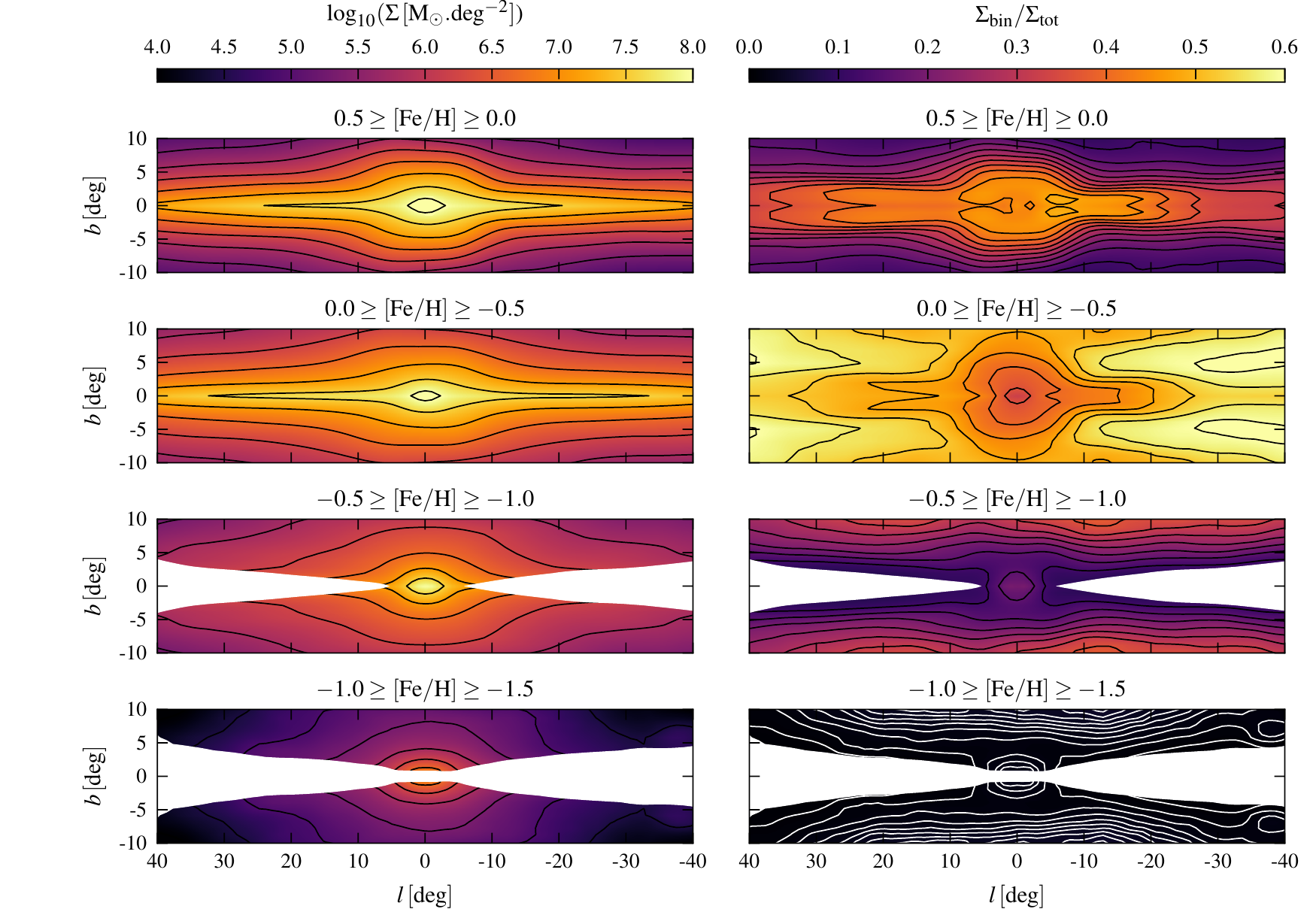}\\
  \caption{Left column: projections in galactic coordinates of the surface density of the fiducial model in the four metallicity bins obtained after fitting the \argos and \apogee metallicity data. Right column: fractional surface density of each metallicity bin. Regions where systematic uncertainties are larger than $30\%$ are masked. The white contour-levels in the bottom right panel are 10 times smaller than in the other panels, as required to highlight the shape of stars in bin D despite their low relative contribution to the total surface density.}
  \label{fig:solarProjections}
\end{figure*}
Another perspective on the spatial distribution of the different metallicities in the Galaxy is given in the galactic coordinates projections of \autoref{fig:solarProjections}. Stars in A and B are found to dominate in the bulge and close to the plane and the distribution of stars in B is found to resemble a thicker replica of the distribution of stars in A. This figure clearly shows the vertical thickening of the stellar density with decreasing metallicity, resulting in a vertical metallicity gradient as discussed in the next subsection.

\subsection{Metallicity gradients in the bar region}
\label{section:metallicityGradients}
The presence of a vertical metallicity gradient along the minor axis of the galactic bulge was first reported by \citet{Minniti1995}. They derived the metallicity of many K giants at $1.5-1.7\kpc$ from the galactic plane and, by comparing to previous metallicity measurements in Baade's window, concluded the presence of a vertical gradient estimated to $-0.35 \, \dex.\kpc^{-1}$. Later studies confirmed the presence of a gradient although its magnitude has a large scatter in the literature. \citet{Zoccali2008} observed a gradient of $ -0.6 \, \dex.\kpc^{-1}$ from bulge K giants between $b=-4\degree$ and $b = -6\degree$, while \citet{Ness2013a} finds a gradient of $ -0.45 \, \dex.\kpc^{-1}$ from 2000 \argos stars between $b=-5\degree$ and $b = -10\degree$. In addition, \citet{Rich2007} found evidence for a flattening of the metallicity gradient between $b=-1\degree$ and $b=-4\degree$. This flattening is also well recovered by the first continuous metallicity maps of the bulge of \citet{Gonzalez2011, Gonzalez2013} who derived photometric metallicities of red giants across the entire bulge from \twomass and \vvv, finding an overall gradient of $-0.28 \, \dex.\kpc^{-1}$ for latitudes between $b=-2\degree$ and $b=-10\degree$. This flattening of the metallicity gradient for $|b|\leq4\degree$ is probably at the origin of the large scatter of the measurements found in the literature, which all refer to different latitude ranges.

\begin{figure*}
  \centering
  \includegraphics{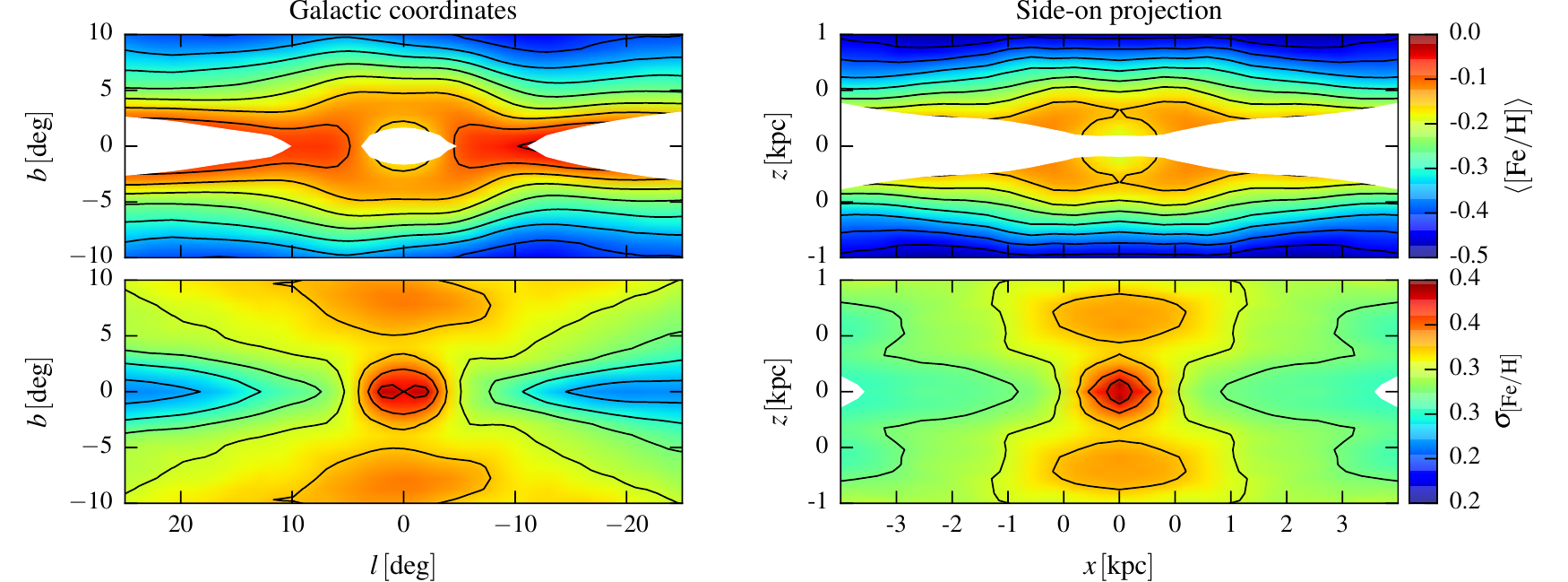}\\
  \caption{Mean metallicity (top) and its standard deviation (bottom) in galactic projection (left) and side-on projection (right). Regions where systematic uncertainties are larger than $0.02 \, \dex$ ($\sim 5\%$) are masked.}
  \label{fig:metallcityMaps}
\end{figure*}
In \autoref{fig:metallcityMaps} we show maps of the mean metallicity and metallicity dispersion of our fiducial model in both galactic and side-on projections. In the plane for  $|b|<5\degree$, the mean metallicity of the model is quite uncertain since in addition to having a broad MDF with a dispersion of $\sim 0.35\, \dex$, it is only constrained by the \apogee data which struggle to reach the bulge. Outside $|b| = 5\degree$ the presence of a vertical gradient is very clear. Along the minor axis of the bulge, including only stars between $6\kpc$ and $10\kpc$ along the line of sight, we find a vertical gradient of the median metallicity in the bulge between $|b| = 5\degree$ and $|b| = 10\degree$ of $-0.45\, \dex.\kpc^{-1}$. This is in good agreement with \citet{Ness2013a}, which is expected since the model fits the \argos data well. This gradient is however larger than the photometric gradient measured by \citet{Gonzalez2013} although the latitude range in the latter study includes part of the strip at $|b|\leq 4\degree$ where the metallicity gradient flattens. Note that the gradient of the mean metallicity differs from that of the median; for our fiducial model, the former is only $-0.33\, \dex.\kpc^{-1}$ in the same region of the bulge. 

Interestingly, the vertical gradient of the median metallicity in the bulge between $|b| = 5\degree$ and $|b| = 10\degree$ of $-0.45\, \dex.\kpc^{-1}$ drops only to $-0.37\, \dex.\kpc^{-1}$ when limited to stars with $\Fe\geq -0.5$. Hence, the vertical metallicity gradient in the bulge is mostly produced by a gradual change in the metallicity of metal-rich stars (i.e stars in A and B), and less by a gradual change in the relative amplitude of a unique metal-rich B/P bulge (stars in A and B together) to a more metal-poor surrounding (stars in C and D together) such as a thick disk \citep{Bekki2011} or a classical bulge \citep{Hill2011}. This is also illustrated by \autoref{fig:metallcityMaps} in which the shape of the B/P bulge, supported by metal-rich stars, remains visible at all heights.

Several interpretations for the origin of this gradient can still be given, from the presence of physically distinct stellar populations \citep{Ness2013a, Fragkoudi2017}, from an initial radial gradient in the early disk \citep{Martinez-Valpuesta2013, DiMatteo2014} or from a continuum of disk populations separated by the bar formation \citep{Debattista2016}.

Outside the bulge, the mean metallicity only slowly changes with radius, with slight signs of flaring for galactocentric distances larger than $3-4\kpc$, as also found by \citet{Bovy2016} who showed from \apogee RCGs that the scaleheight of more metal-poor stars increases with increasing radius.

\subsection{A central concentration of metal-poor stars: classical bulge or inner stellar halo}
\label{section:centralConcentrationCandD}

\begin{figure}
  \centering
  \includegraphics{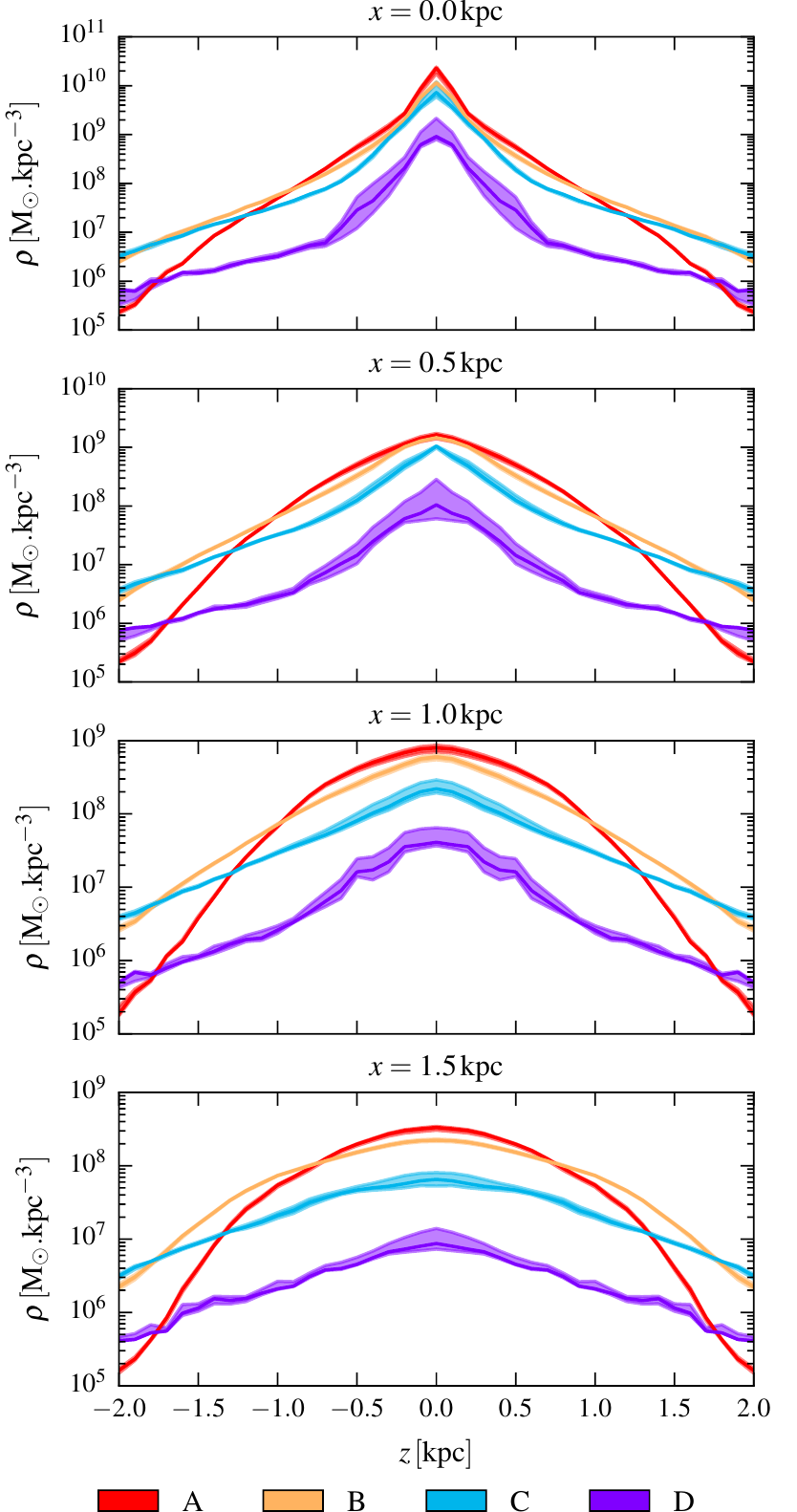}\\
  \caption{Vertical density profiles in the four metallicity bins for cuts along the major axis of the bar, at $x=0, 0.5, 1.0$ and $1.5\kpc$ (from top to bottom respectively) in the coordinates of \autoref{fig:SurfaceDensityBestModel} for our fiducial model (solid lines) and range of variation models (coloured areas). In C and D the vertical densities have similar profiles that clearly show the presence of an inner and centrally concentrated metal-poor structure, possibly a classical bulge or the inner stellar halo.}
  \label{fig:verticalprofiles}
\end{figure}

\autoref{fig:allSurfaceDensitiesOfComponents} shows that the stellar densities in both bins C and D are more centrally concentrated in the inner $\kpc$ of the Galaxy than they are on larger scales. This is shown in more detail in \autoref{fig:verticalprofiles} where we plot the vertical density at different points along the major axis of the bar for all four metallicity bins. 
In this figure, stars in bins C and D have similar vertical density profiles for all cuts. For cuts located at major axis distance $x$ larger than a $\kpc$, they share the vertical exponential profile of a thick component with scaleheight $\sim 500\pc$. However, in the inner $\kpc$ of the Galaxy, an extra component with short scaleheight rises towards the Galactic Centre and is seen consistently only in the metal-poor bins C and D. For the central cut, the scaleheight of this inner metal-poor component is only $\sim 70\pc$, and it vanishes below the large-scale thick disk density at $|z|\geq 600\pc$. 

The presence of a central concentration of metal-poor stars in the inner $\kpc$ of the Galaxy has already been reported by several authors using different data sets. Both \citet{Pietrukowicz2015} and \citet{Dekany2013} found, respectively from \ogle and \vvv data, that RRLyrae stars in the bulge were centrally concentrated, although they disagree about the shape of the RRLyrae density distribution. In addition, \citet{Schultheis2015} reported the presence of a peak of metal-poor stars in the \apogee data within the central degree of the Galaxy. Finally, \citet{Zoccali2016} combined spectroscopic data from the recent \gibs survey with the bulge stellar density from \citet{Valenti2015} and find a central concentration of metal-poor stars at $b=-2\degree$ that vanishes at $b=-4\degree$. 

Interpreting the 3D distribution and central concentration of stars in bins C and D in terms of physical components requires more data and is beyond the scope of this paper. Theoretically at least three metal-poor components are expected or suspected to contribute to the metallicity bins C and D in the inner Galaxy. The thick disk, locally metal-poor with a flat radial metallicity gradient \citep{Bensby2007, Cheng2011, Bovy2015b} is expected to contribute to both C and D in the inner Galaxy, and could possibly be centrally compressed because of the deep gravitational potential of the nuclear bulge (\citealt{Launhardt2002}; \hyperlink{P17}{P17}). In addition, the stellar halo with its steep radial profile \citep{Wetterer1996,Juric2008} is also a possible candidate \citep{Perez-Villegas2016}, although it is not clear that its mass in the bulge would be sufficient to provide a significant contribution to C and D. Finally, the central concentration of metal-poor stars could also indicate the presence of a classical bulge component, a fossil of early mergers but whose existence has not yet been established. 

To conclude, from the 3D mass distributions of stars with metallicity in the bins C and D obtained by our modelling, it seems that on large-scale bin C stars are mainly distributed as a thick disk. This is supported by its flattening of only $\sim0.3$, quite extreme for a significant contribution from a classical bulge or from the inner stellar halo, according to both observations of external galaxies \citep{Kormendy2004} and simulations (flattening $\sim 0.5$ for low-mass classical bulge in \citealt{Saha2011}; or $\sim 0.6$ for the inner stellar halo in \citealt{Perez-Villegas2016}). Inside the central $\kpc$, an extra metal-poor component dominates but a number of interpretations are possible. More data would be required to investigate this further.

\subsection{Support to the B/P bulge and bar}
\label{section:supportBar}

\begin{figure}
  \centering
  \includegraphics{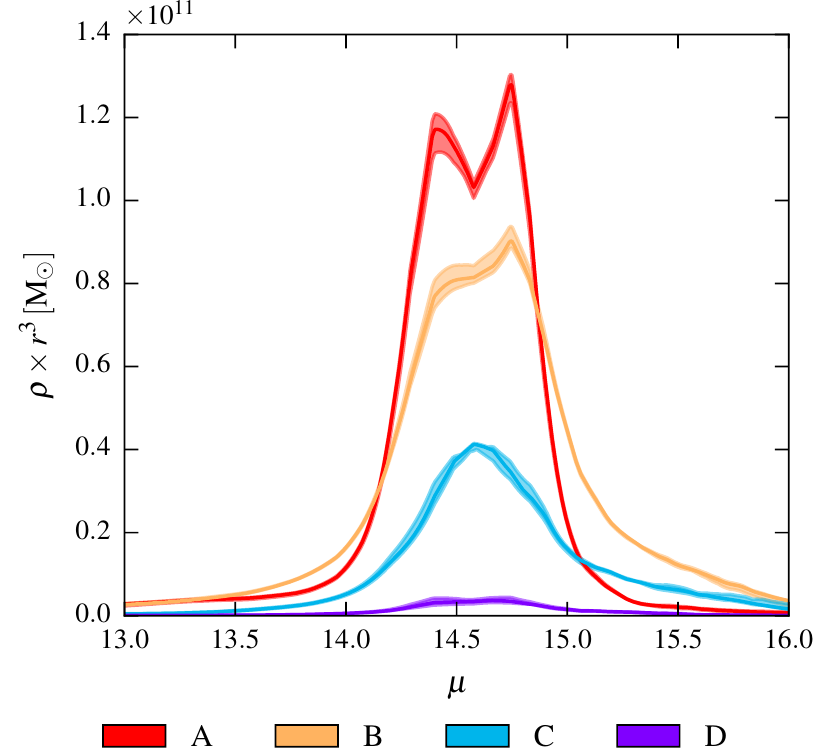}\\
  \caption{`Split red clump' at different metallicities in a minor axis field at $b=-5\degree$ for the fiducial model (solid lines) and range of variation models (coloured areas). The plotted quantity is $\rho\times r^3$ as a function of distance modulus, i.e. the equivalent of the number count of stars as a function of distance modulus for a perfect standard candle stellar population. The `split red clump' is only visible in metallicity bins A and B.}
  \label{fig:splitRedClump}
\end{figure}
In the \argos bulge sample, only stars in A and B contribute to the B/P shape. This is demonstrated by \autoref{fig:splitRedClump} where we show the stellar density times the cube of the line-of-sight distance ($\rho\times r^3$), as a function of distance modulus for the four metallicity bins in a minor axis field at $b=-5\degree$. The quantity plotted in this figure $\rho\times r^3$ is proportional to the number count of stars as a function of distance modulus for a standard candle stellar population. This figure can thus be compared to the similar diagram of \citet{Ness2012} where they show from the raw \argos data directly that only stars in A and B have a bimodal magnitude distribution, the so-called `split red clump', and that therefore only stars in A and B support the B/P shape of the bulge. Our model confirms these results, taking also into account the \argos selection function.

\begin{figure}
  \centering
  \includegraphics{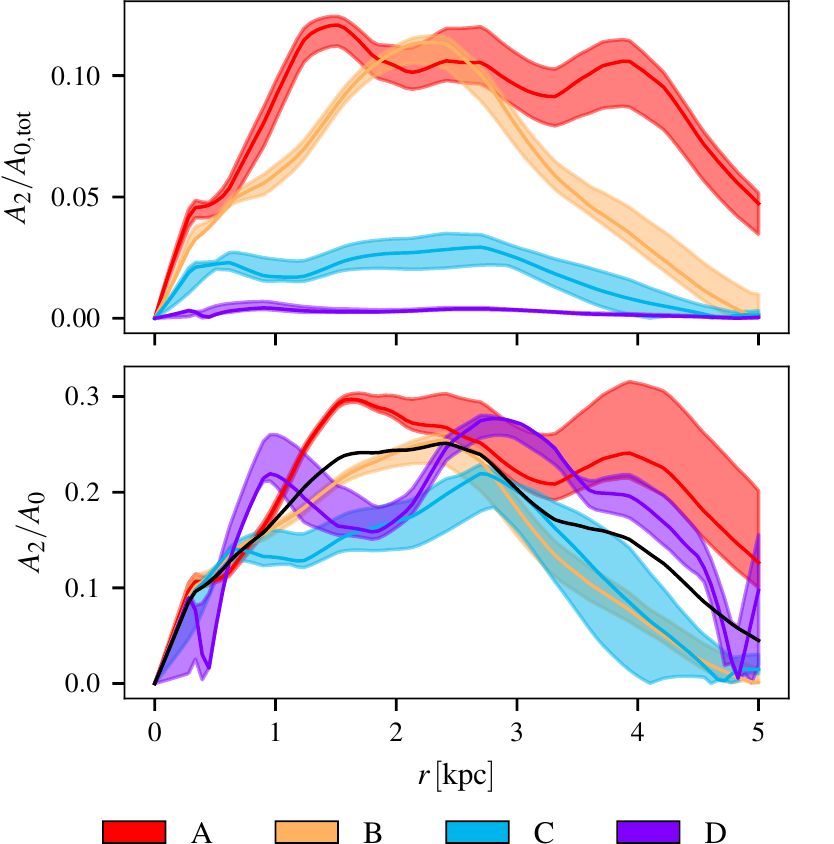}\\
  \caption{Radial profiles of the $m=2$ Fourier coefficients in the four metallicity bins relative to the overall $m=0$ Fourier coefficient (upper panel) and relative to the $m=0$ Fourier coefficient of the considered metallicity bin (lower panel). The solid lines refers to our fiducial model and coloured areas to the range of variation models. The black line in the lower panel shows the relative $m=2$ Fourier coefficient for the total mass, i.e. all metallicity bins combined.}
  \label{fig:barstrength}
\end{figure}
From the face-on views of \autoref{fig:allSurfaceDensitiesOfComponents} it is clear that stars of all metallicities support the bar, but in terms of mass, most of the support to the bar is provided by metal-rich stars.
 This is shown by the upper panel of \autoref{fig:barstrength} where we plot the radial profiles of the $m=2$ Fourier coefficients in the four metallicity bins relative to the overall $m=0$ Fourier coefficient. Metal-poor stars in bins C and D provide only a weak mass support to the bar, mostly because they represent only a small fraction of the mass overall (see \autoref{table:masses}). However, there is still a significant fraction of the metal-poor stars that does support the bar, as shown by the lower panel of \autoref{fig:barstrength} where we plot the radial profiles of the $m=2$ Fourier coefficients in the four metallicity bins relative to the $m=0$ Fourier coefficients of the respective metallicity bin.

 Stars that do not strongly support the bar are instead building the inner disk. To isolate the bar from the inner disk we classify the N-body orbits of the model of \hyperlink{P17}{P17} into two classes, the bar-supporting orbits and the not bar-supporting orbits. Following the method of \citet{Portail2015b}, we identify the bar-supporting orbits as the orbits for which $f_r/f_x = 2 \pm0.1$, where $f_r$ and $f_x$ are respectively the dominant frequency of the oscillations in cylindrical radius and in the bar major axis coordinate along the orbits. 
\begin{figure}
  \centering
  \includegraphics{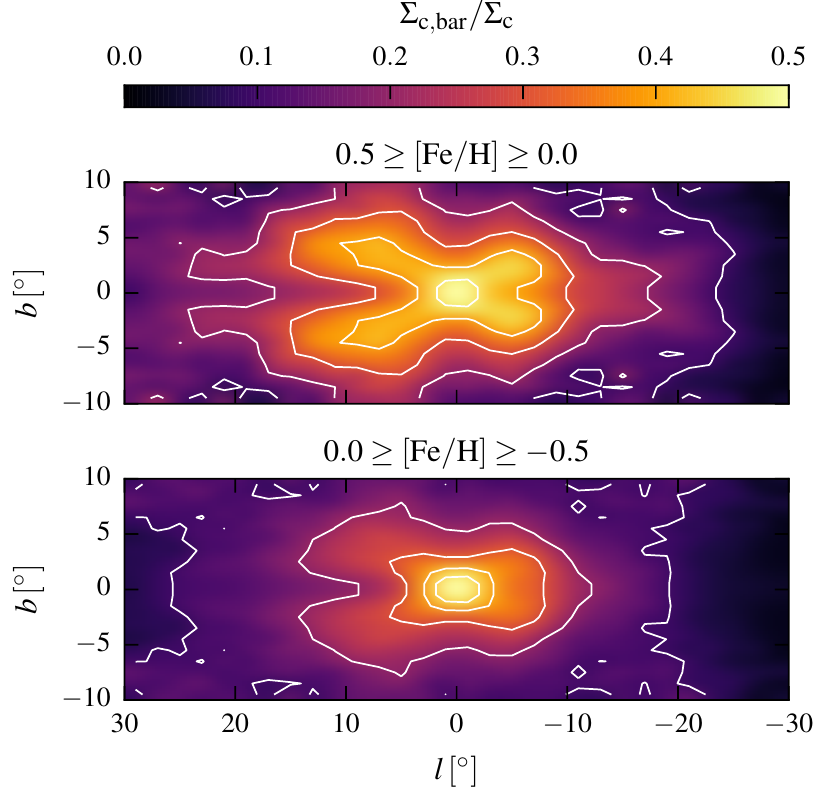}\\
  \caption{Fraction of the surface density of stars in bin A (top) and bin B (bottom) in galactic coordinates that originates from bar-supporting orbits. The support of stars of bin A to the B/P shape of the bulge is beautifully striking.}
  \label{fig:barSupportingFractions}
\end{figure}
In \autoref{fig:barSupportingFractions}, we show for bins A and B the fraction of the surface density in galactic coordinates that originates from bar-supporting orbits. Both metallicity bins exhibit an increase of bar-supporting orbits towards the corners of the bulge, as expected for orbits supporting the B/P shape of the bulge. Note that for bin A about half the stars in this ``orbital X-shape'' support the bar.

How are the bar and inner disk orbits divided between the different metallicity bins? Using our orbit classification, we find that the galactic bar is composed of $52\% \pm 3\%$ of stars with metallicities within bin A, $34\% \pm 1\%$ of stars with metallicities within bin B, and only $12\% \pm 1\%$ and $2\% \pm 1\%$ of stars with metallicities within bin C and D, respectively. For the inner disk within corotation, the fractions of stars in A, B, C and D are respectively $38\% \pm 3\%$, $47\% \pm2\%$, $14\% \pm1\%$ and $1\% \pm 1\%$. Note that although B is found in \citet{Ness2013b} to be the most populated metallicity bin in the \argos data, we find here that bin A stars give the main mass support to the bar and have a similar role as bin B stars in building the inner disk.

\section{Kinematic distribution of the metallicity in the Galaxy}
\label{section:kinematicVariations}
In this section, we show how stars in the different metallicity bins differ by their kinematics and orbital properties. As already pointed by \citet{Babusiaux2010} and \citet{Ness2013b}, metal-rich stars with $\Fe\geq-0.5$ (i.e, in bins A and B) have very different kinematics than more metal-poor stars with $\Fe \leq-0.5$ (i.e in bins C and D). We now analyse in detail these two main groups separately using our fiducial chemodynamical model of \autoref{section:fiducialAndVariations}.

\subsection{Bar-like kinematics of metal-rich stars}
The kinematics of stars in bin A and B are found in the \argos data to be similar to each other, as shown by \citet{Ness2013b}. They are both rapidly and cylindrically rotating, i.e with only weak variations as a function of latitude. They have similar dispersion profiles, both steeply rising towards the centre. In the \argos data, bin B stars are found to rotate slightly faster than bin A stars, with a velocity dispersion about $20\kms$ higher at the centre. 

The similarities between the kinematics of stars in these two metallicity bins have been interpreted as an indication of a common thin disk origin by \citet{DiMatteo2014}. In their N-body simulations of bar forming disks, stars that formed on average at larger radii are mapped after bar formation and buckling into a population that exhibits faster rotation and higher dispersion. Thus, with an initially decreasing metallicity gradient in the disk, birth radius and metallicity become correlated and naturally reproduce the kinematic behaviour of bin A and B stars, thus arguing for a common disk origin for these stars.

This simple picture however does not entirely capture the trend we see in our fiducial chemodynamical model. 
\begin{figure*}
  \centering
  \includegraphics{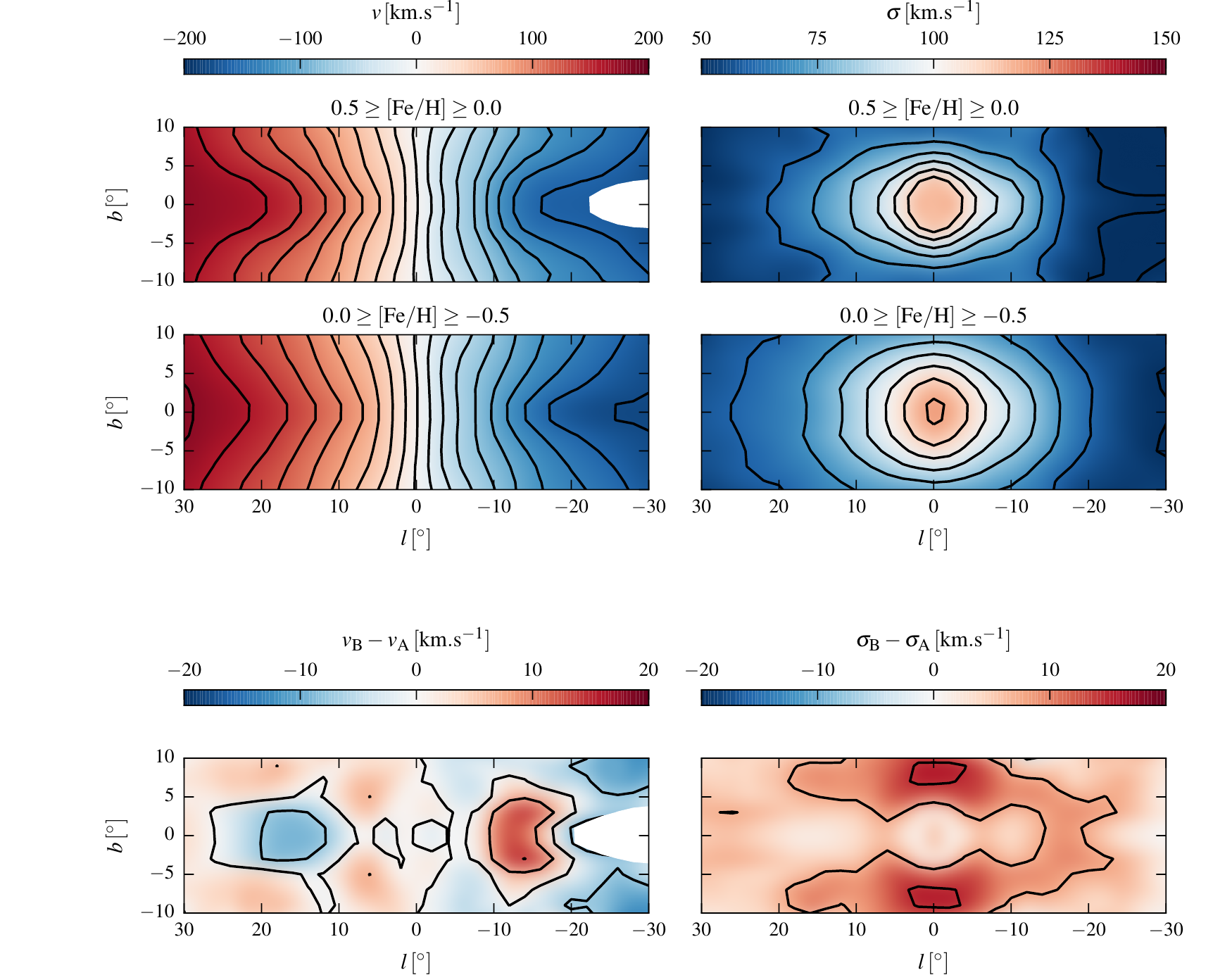}\\
  \caption{Upper four panels: Maps of the mean radial velocity (left) and velocity dispersion (right) for stars in bin A (top) and bin B (bottom). The bottom two panels show the difference maps between B and A. Stars in B are hotter and rotate faster than stars in A for $|b|\geq 5\degree$, as reported in \citet{Ness2013b}. We find however that for $|b|\leq 5\degree$, stars in B appear to rotate slower than those in A. Regions where systematic uncertainties are larger than $2\kms$ are masked.}
  \label{fig:KinematicsOfAandB}
\end{figure*}
\autoref{fig:KinematicsOfAandB} shows maps in galactic coordinates of the mean radial velocity and velocity dispersion for stars in bins A and B, integrating along the line of sight between $3$ and $12\kpc$ from the Sun. We see that B is indeed faster than A for $b\geq5\degree$ where the \argos fields are located, but not in the plane for $b\leq5\degree$ where we find the inverse trend, with A being about $10\kms$ faster than B at $l=20\degree$. This result likely originates from the complex orbital structure of the in-plane bar, mostly populated by bin A stars. Although the study of the orbital structure of the bulge and bar in the dynamical model of \hyperlink{P17}{P17} is beyond the scope of this paper, we can investigate the origin of this separate relative behaviour of A and B in the plane by experimenting with various cuts in the model. Doing so we find that the particles that are responsible for A being faster than B in the plane are located less than $2\kpc$ away from the bar major axis, and predominantly along the rear-side of the bar. These particles produce a high-velocity tail in the line-of-sight distributions, thus shifting the mean velocity in bin A to higher values. We thus speculate that the faster rotation of in-plane stars in bin A is likely to be caused by the complex orbital structure of the bar, in possible relation with the orbits suggested by \citet{Aumer2015} to be at the origin of the high-velocity peak observed in the \apogee data for $l = 4 \degree - 14\degree$ and $|b| \leq 2 \degree$ \citep{Nidever2012}. Note however that the model only shows a high-velocity tail, with no sign for a separate high-velocity peak. It appears only for $l\geq 8\degree$ and is thus unrelated to the possible presence of a $\sim \kpc$ scale nuclear disk as suggested by \citet{Debattista2015}. A detailed orbit analysis would be required to investigate the cause of the fast rotation of bin A stars further.

\subsection{Kinematics of the metal-poor stars}
The metal-poor stars, often defined by $\Fe\leq-0.5$, have been shown to be kinematically distinct from their metal-rich counterparts. They generally rotate more slowly and have a higher dispersion, showing that metal-poor stars cannot be seen as just the metal-poor tail of a single population with a broad MDF. 
\begin{figure*}
  \centering
  \includegraphics{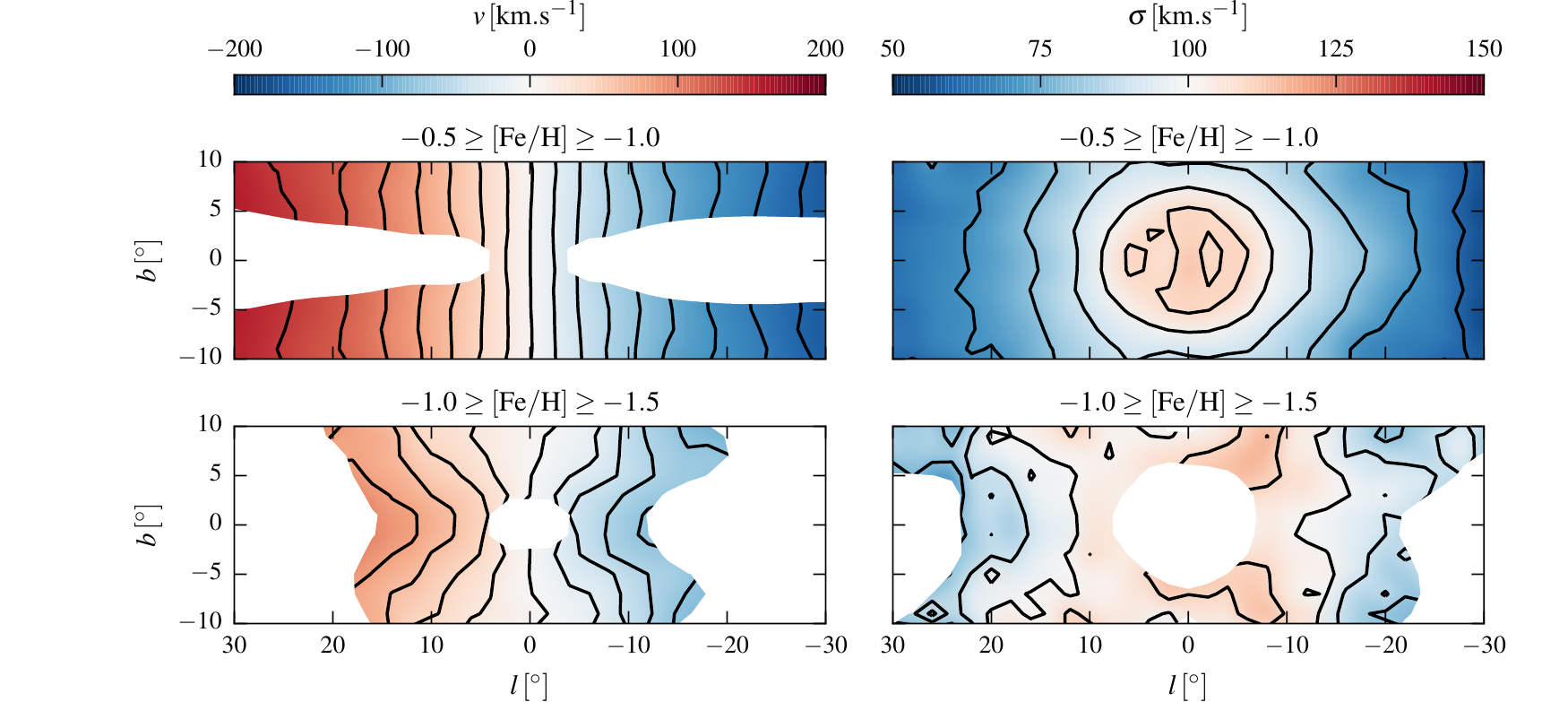}\\
  \caption{Mean radial velocity (left) and velocity dispersion (right) of stars in the metallicity bins C (top) and D (bottom). Regions where systematic uncertainties are larger than $2\kms$ are masked.}
  \label{fig:KinematicsOfCandD}
\end{figure*}
This is illustrated in more detail in \autoref{fig:KinematicsOfCandD} where we show maps of the mean radial velocity and velocity dispersion of stars in bins C and D from our fiducial model. By comparing with \autoref{fig:KinematicsOfAandB}, both C and D appear to have a higher dispersion than their surroundings, except in the central few degrees where the metal-rich stars have larger dispersions, caused by their bar-supporting orbits. \citet{Debattista2016} showed recently that hotter stellar populations are less responsive to the bar formation and buckling. Similar conclusion is obtained by \citet{DiMatteo2016} for vertically thicker stellar populations. Thus if C and D were initially populated by a hot and/or thick stellar population, this would explain their relative weak support to the bar and their absence of bar-like kinematic features. We note however that metal-poor stars with $\Fe\leq-0.5$ do not share a unique kinematic behaviour but rather show smoothly evolving trends with metallicity. In the \argos data, bin C stars at $b=-5\degree$ reach a mean velocity of $\sim 120\kms$ at $l=20\degree$ while bin D stars reach only $\sim 70\kms$. 

\begin{figure}
  \centering
  \includegraphics{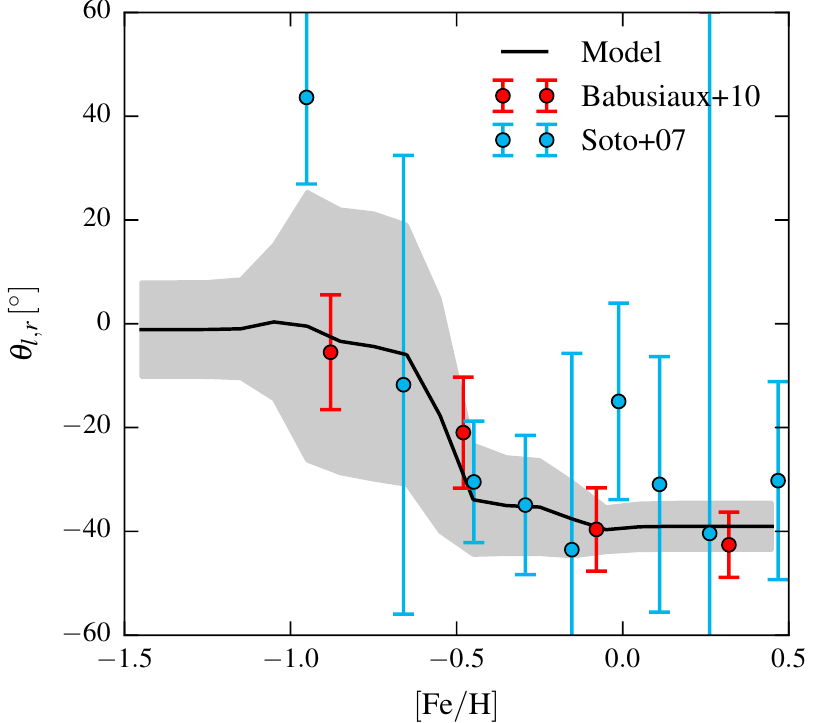}\\
  \caption{Vertex deviation $\theta_{l,r}$ in Baade's window as a function of metallicity for our fiducial model compared to the data of \citet{Babusiaux2010} and \citet{Soto2006}. In computing the model predictions we include only stellar particles between $6\kpc$ and $10\kpc$ away along the line of sight. The shaded area indicates the $1\sigma$ statistical error on the model vertex deviation. The large errors in the model predictions at low metallicity arise from the fact that $\sigma_{r}^2$ and $\sigma_{l}^2$ are increasingly similar to each other with decreasing $\Fe$, and therefore the angle of the near-circular velocity ellipsoid is poorly defined.}
  \label{fig:VertexDeviation}
\end{figure}

\subsection{Vertex deviation in the bulge}
\label{section:vertexdeviation}
Because the physical components present in the inner Galaxy (at least thick disk, stellar halo and possibly classical bulge) are likely to overlap in metallicity space, we expect smooth kinematic transitions as  a function of metallicity. This is illustrated by \autoref{fig:VertexDeviation} where we show the vertex deviation in Baade's window $\theta_{r,l}$ of our model compared to the data from \citet{Soto2006} and \citet{Babusiaux2010}. The vertex deviation $\theta_{r,l}$ is defined by 
\begin{equation}
 \tan{(2\theta_{r,l})} = \frac{2 \sigma^2_{r,l}}{|\sigma_{r}^2 - \sigma^2_{l}|}
\end{equation}
where $\sigma^2_{r,l}$, $\sigma_{r}^2$ and $\sigma_{l}^2$ are respectively the covariances of the radial velocity and proper motion dispersion in the longitudinal direction, and their two associated variances. The vertex deviation represents the angle of the principal axis of the velocity ellipsoid in non-isotropic cases. We selected only stellar particles between $6\kpc$ and $10\kpc$ away along the line of sight within a cone of half a degree centred on $l = 1\degree$, $b = -3.9\degree$ and evaluated the vertex deviation of the model as a function of metallicity on a grid with a resolution of $0.1\dex$.

It is clear in \autoref{fig:VertexDeviation} that $\theta_{r,l}$ is significantly non-zero for stars with $\Fe\geq-0.5$ and gradually transitions to zero for more metal-poor stars. Our fiducial model reproduces the observed evolution of $\theta_{r,l}$ with metallicity well, providing a good fit to the data although not directly fitted to it.

The absence of a significant vertex deviation for $-1 \leq \Fe \leq -0.5$ (C) has been interpreted by \citet{Soto2006} and \citet{Babusiaux2010} as a signature of the presence of an old metal-poor spheroid. Our model shows instead that bin C stars have no significant vertex deviation $\theta_{r,l}$ in Baade's window, which is shown in \autoref{fig:verticalprofiles} to be sufficiently far away from the Galactic Centre to be dominated there by the large-scale disk-like density of bin C stars found in \autoref{section:spatialProjections}. In addition, C is shown in \autoref{fig:KinematicsOfCandD} to be cylindrically rotating with a rotation only slightly slower than that of the metal-rich stars. 

We thus conclude that our interpretation for stars in the metallicity bin C forming a thick disk outside the central $\kpc$ is consistent with its hot kinematics and relatively fast rotation. The lack of significant vertex deviation for metal-poor stars in Baade's window is caused by the large-scale thick disk distribution of bin C stars, with no need for a large classical bulge component.

Little can be reliably said from our model about the metallicity bin D since it is only poorly constrained by the data. D is found in the model to be cylindrically rotating, but at a slower rate than C and has a dispersion that varies only slightly in the inner $10\degree$ of the Galaxy. Differences between C and D are likely to highlight different contributions of the underlying metal-poor stellar populations described above. For example, the lower rotation of D with respect to C could be explained by a higher contribution in D of a non-rotating stellar halo.

\section{Discussion and conclusion}
\label{section:discussion}

We presented an extension of the Made-to-Measure (M2M) method for modelling self-consistent equilibrium chemodynamical models of galaxies. Its application in the inner Milky Way results in the first chemodynamical particle model fitted to chemo-kinematic data in the bulge and bar. Such chemodynamical models will be invaluable in the near-future for interpreting chemical abundances and stellar parameters provided by the many on-going and future large-scale spectroscopic surveys of the Galaxy. Indeed, M2M chemodynamical models allow to maximize the information extracted from spectroscopic data by constraining at once the spatial distribution and kinematics of stars with the orbital distribution of a self-consistent dynamical model. Since it is built upon the standard M2M method, it is flexible enough to be applicable to complex systems such as the inner barred galactic disk, where other modelling technique would be either not applicable or too complex. We applied the method based on the dynamical model of the bar region recently made by \citet{Portail2016a}, and successfully reproduced the spatial and kinematic variations of the metallicities of stars seen in the \argos and \apogee (DR12) surveys. Our chemodynamical model allows us to study the dynamics of stars with different metallicities in great detail, unveiling their distinct 3D density distributions, kinematics and orbital structures. Summarizing the properties of stars in the different metallicity bins in the inner $5\kpc$ of the Galaxy, as found in our model, but also in the light of the work of \citet{Ness2013a, Ness2013b} and \citet{DiMatteo2014, DiMatteo2015}, we conclude the following: 

\begin{description}
 \item[ A ($0.5 \geq \Fe > 0$):] We find that A is the metallicity bin that contributes most to the galactic bar and bulge structure in its current state. A is strongly barred and hosts $52\% \pm 3\%$ of the bar-supporting orbits. It is generally thin, dominates over the other metallicity bins in the plane and provides a very strong support to the B/P shape of the bulge. Its hammer-like shape in the outer parts of the bar indicates that a significant fraction of A belongs to the superthin bar. A has a very low dispersion of only about $\sigma \approx 60-80\kms$ that steeply rises in the central $5\degree$, caused by the bar-supporting orbits. Stars in A are found to be more rapidly rotating than in the other metallicity bins for longitudes between $10$ and $30\degree$ at $|b| \leq 5\degree$ and show a high-velocity tail in the line-of-sight velocity distribution there. We suggest that this tail, shifting the mean velocity to a higher value, is caused by the complex orbital structure of the bar in the plane. A also contributes in our model $38\% \pm 3\%$ of the inner disk within corotation. In good agreement with \citet{Ness2013b} and \citet{DiMatteo2015}, bin A is thus consistent with being partly populated by stars with an early thin disk origin. However, since the superthin bar is probably associated with a more recent star formation episode \citep{Wegg2015}, bin A stars are probably a composite stellar population, with old and metal-rich stars formed in the early disk and later mapped into the bar, together with stars formed more recently in the superthin bar or trapped by the bar.
  
 \item[ B ($0 \geq \Fe > -0.5$):] We find that B is the second main metallicity bin of the galactic bulge and bar structure, which dominates over A at $|b| \geq 5\degree$. B is also strongly barred, contains $34\% \pm 1\%$ of the bar-supporting orbits and also contributes to the B/P shape of the bulge. B is found to be more extended vertically than A, in good agreement with \citet{Ness2013a}. It is generally hotter than A and also shows a steep rise in the velocity dispersion towards the centre, a signature of its barred nature. B rotates generally faster than A at $|b|\geq 5\degree$, except in the plane where it does not share the complex orbital structure of A. It accounts for $47\% \pm2\%$ of the inner disk, a similar share to that of bin A. B thus appears to be consistent with being populated by stars with a disk origin, formed from stars initially located at larger radii than bin A stars, as found by \citet{DiMatteo2014} to explain the faster rotation and larger dispersion.
 
 \item[ C ($-0.5\geq \Fe > -1$):] C shows significant differences from A and B. It is thicker, and has a slightly slower rotation and a slightly higher dispersion than the more meal-rich stars. C is found to contribute weaker support to the bar, hosting only $12\% \pm 1\%$ of the bar-supporting orbits. Outside the central $\kpc$, C is strongly vertically flattened ($\sim 0.3$) and has a thick disk like density with a fast and cylindrical rotation, while in the inner $\kpc$, it becomes significantly more centrally concentrated. Several metal-poor components are expected to be present in the inner Galaxy, thick disk, stellar halo and possibly a classical bulge which would all be expected to contribute to the metallicity bin C at some level. From both its spatial distribution and kinematics we conclude that outside the central $\kpc$ C is mainly populated by thick disk stars. Whether or not bin C stars were already in a thick disk before bar formation, as suggested by \citet{DiMatteo2015}, is however still unclear. \citet{Debattista2016} showed instead that a thin but initially radially hot disk population could also be mapped into a thick distribution after the bar formation. In the central $\kpc$ of the Galaxy, the concentration of metal-poor stars can be interpreted in various ways and more data would be required to investigate this further.
  
 \item[ D ($-1 \geq \Fe > -1.5$):] D is only weakly constrained by the \argos data, and even less by the \apogee data since the fraction of metal-poor stars increases with height above the plane. In our models, D is found to host a very thick, high dispersion and slowly rotating stellar population. It is also centrally concentrated in the inner $\kpc$, similarly to C. However, D is different from a metal-poor tail of C, with a rotation that is not cylindrical and a nearly constant dispersion in the inner $10\degree$ of the Galaxy, possibly indicating the contribution of an extended non-rotating and metal-poor component which we speculate could be the extended stellar halo.
\end{description}

To conclude, most of the metal-rich stars (B and most of A) seem to be consistent with a common and relatively simple origin, the early thin disk, which after bar formation and buckling became mapped into different spatial distributions and a rather complex orbital structure. Part of A could also be associated to a more recent star formation episode, as indicated by the similarities between the superthin bar and the shape of the bar end seen in this metallicity bin. On the contrary, metal-poor stars (C and D stars) appear to presently have a relatively simple structure, only weakly bar-following but they are probably composed of a complex superposition of components: thick disk (either originally thick or thickened by the bar formation), inner halo and possibly a classical bulge. Note however that in our model, the lack of significant vertex deviation for metal-poor stars in Baade's window is naturally reproduced by the thick disk distribution of bin C stars and therefore does not argue for the presence of a classical bulge in the Milky Way.

Our M2M chemodynamical model of the bar region, although the first of its kind, represents a significant step towards more complete models, that by including additional chemical dimensions will be able to better constrain the formation history of the Galaxy. On-going and future spectroscopic surveys such as APOGEE, GAIA-ESO, GALAH and MOONS will provide stellar abundances of 10-20 elements for more than $10^5$ stars each. This profusion of data allows extending the dimensionality of the chemical phase-space in M2M chemodynamical models to other elements that better constrain the origin of stars. The abundance in $\alpha$-elements is of primary importance, since it is a natural chemical ``clock'' giving the time-scale of stars formation of a given stellar population. We plan in the near future to first extend the modelling of the bar region to a two dimensional chemical space, including $\Fe$ and $\rm{[\alpha/Fe]}$, and later to other elements and stellar ages.

\section*{Acknowledgments}
We acknowledge the use of the p3lib potential solver, made available to us by Jerry Sellwood and Monica Valluri. We are grateful to Jo Bovy for publicly releasing his tools dealing with \apogee data, and particularly its selection function. We thank the organizers of the conferences ``The Milky Way and its environment'' in Paris and GASP 2016 in Canberra for the opportunity to present this work before publication. We thank the anonymous referee for precise reading and helpful comments.


\phantomsection\label{lastpage}
\end{document}